%% file: 5PNv2.tex
\documentclass[11pt]{article}
\pdfoutput=1

\usepackage{jheppub}
\usepackage[utf8]{inputenc}
\usepackage{booktabs}
\usepackage{graphicx}
\usepackage{bbold}
\usepackage{subcaption}
\usepackage{mathtools}
\usepackage{mathrsfs}
\usepackage{amsmath}
\usepackage{kantlipsum}
\usepackage{comment}
\allowdisplaybreaks
\usepackage{autobreak}

\numberwithin{equation}{section}

\newcommand{\nn}{\nonumber}

\newcommand\beq{\begin{equation}}
\newcommand\eeq{\end{equation}}
\newcommand\beal{\begin{aligned}}
\newcommand\eeal{\end{aligned}}
\newcommand\bea{\begin{eqnarray}}
\newcommand\eea{\end{eqnarray}}

\newcommand\dd{{\mathrm d}}
\usepackage{xcolor}

\newcommand{\bd}{{\boldsymbol d}}
\newcommand{\bb}{{\boldsymbol b}}
\newcommand{\bk}{{\boldsymbol k}}

\newcommand{\bn}{{\boldsymbol n}}
\newcommand{\bp}{{\boldsymbol p}}

\newcommand{\bL}{{\boldsymbol L}}
\newcommand{\bA}{{\boldsymbol A}}

\newcommand{\bq}{{\boldsymbol q}}
\newcommand{\br}{{\boldsymbol r}}

\newcommand{\bx}{{\boldsymbol x}}
\newcommand{\bv}{{\boldsymbol v}}

\newcommand{\ba}{{\boldsymbol a}}
\newcommand{\bE}{{\boldsymbol E}}

\newcommand{\bF}{{\boldsymbol F}}

\newcommand{\PV}{\mathscr{P}}
\newcommand{\sg}{\textrm{Sign}}

\newcommand{\ord}[1]{^{(#1)}}

\newcommand{\Pp}{\mathbb{P}}

\def\ddl{\delta\!\!\!{}^-}

\makeatletter
\newcommand{\Biggg}{\bBigg@{3.5}}
\makeatother

\input{diags.tex}

\tikzfeynmanset{scalmom/.style = {
   decoration={
     markings,
     mark=at position 0.5
          with {\arrow{stealth[width=1.5mm,length=1.5mm]}}
     },
   postaction=decorate}
}

\tikzfeynmanset{antiscalmom/.style = {
   decoration={
     markings,
     mark=at position 0.5
          with {\arrow{stealth reversed[width=1.5mm,length=1.5mm]}}
     },
   postaction=decorate}
}

\begin{document}
\preprint{DESY\, 24-133\\\phantom{~}}
\title{ \center Nonlinear Gravitational Radiation Reaction:\\ [0.2cm] 
Failed Tail, Memories \& Squares} 
\author{\center \large Rafael A. Porto,}
\author{\large Massimiliano M. Riva,}
\author{and \large Zixin Yang}

 \affiliation{\hspace{3.4cm} Deutsches Elektronen-Synchrotron DESY,\,Germany.}

\abstract{Using the Schwinger-Keldysh ``in-in" effective field theory (EFT) framework, we complete the knowledge of nonlinear gravitational radiation-reaction effects in the (relative) dynamics of binary systems at fifth Post-Newtonian (5PN) order. Diffeomorphism invariance plays a key role guaranteeing that the Ward identities are obeyed (in background-field gauge). Nonlocal-in-time (memory) effects appear in the soft-frequency limit as boundary terms in the effective action, consistently with the loss of (canonical) angular momentum. We identify a {\it conservative} sector through Feynman's $i0^+$-prescription. Notably, terms at second order in the (linear) radiation-reaction force also produce conservative-like effects (as we likewise demonstrate in electromagnetism). For the sake of comparison, we derive the ${\cal O}(G^4)$ contribution to the total (even-in-velocity) 5PN relative scattering angle. We find perfect agreement in the overlap with the state of the art in the Post-Minkowskian expansion, both in the conservative and dissipative sectors, resolving the (apparent) discrepancy with previous EFT results. We will return to the full conservative part of the 5PN dynamics elsewhere.} 
\maketitle

\section{Introduction}

Einstein's theory of gravity is rooted on the nonlinearities of its field equations, famously producing black hole (Schwarzschild and Kerr) solutions in vacuum. Yet, in many situations of interest, such as the dynamics of a pair of black holes emitting the gravitational waves (GWs) observed by the LIGO-Virgo-KAGRA collaboration \cite{KAGRA:2021vkt}, finding even an approximate solution within a perturbative scheme---such as the weak-field/slow-velocity Post-Newtonian (PN), weak-field Post-Minkowskian (PM), or small-mass-ratio expansions---is a daunting task. Fortunately, this is no longer an academic endeavour, since GW astronomy with third-generation detectors such as LISA \cite{amaroseoane2017laser}, the Einstein Telescope \cite{Punturo:2010zz} and the Cosmic Explorer \cite{Reitze:2019iox} relies upon our ability to produce high-precision theoretical models for compact binaries \cite{Buonanno:2014aza,music,AlvesBatista:2021eeu,Bernitt:2022aoa}. Its intricacy, however, requires an orchestrated effort joining various `traditional'~\cite{Blanchet:2013haa,Schafer:2018jfw,Barack:2018yvs},  effective field theory (EFT)~\cite{Goldberger:2004jt,Porto:2005ac,Goldberger:2009qd,Galley:2009px,Galley:2014wla,Porto:2016pyg,Kalin:2019rwq,Kalin:2019inp,Kalin:2020mvi,Mogull:2020sak,Mougiakakos:2021ckm,Dlapa:2021npj,Dlapa:2021vgp,Cho:2021arx,Cho:2022syn,Kalin:2022hph,Dlapa:2022lmu,Jakobsen:2022psy,Goldberger:2022ebt,Dlapa:2023hsl,Dlapa:2024cje,Amalberti:2024jaa}, and amplitudes-based~\cite{Neill:2013wsa,Bjerrum-Bohr:2018xdl,Cheung:2018wkq,Bern:2019crd,Brandhuber:2021eyq,Bern:2021dqo,Bern:2021yeh,DiVecchia:2023frv} analytic methodologies, combined with numerical simulations~\cite{Campanelli:2005dd,Pretorius:2005gq}, to tackle the two-body problem in general relativity.\vskip 4pt 

Among the key contributions to the gravitational dynamics are the so-called {\it hereditary} terms, e.g.~\cite{Blanchet:1992br,Blanchet:1997ji}, which are due to the interaction of the outgoing radiation with the binary's (Kerr) background geometry, a.k.a.~``tail"  and ``failed-tail" effects, as well as the waves emitted at an earlier time, a.k.a.~``memory" effects. These nonlinear gravitational corrections, which are not present in  electromagnetism, not only modify the GW radiated power \cite{Goldberger:2009qd,Blanchet:2023bwj}, they also contribute, starting at 4PN order \cite{Bini:2013zaa,Damour:2014jta,Galley:2015kus,Bernard:2015njp}, to the {\it conservative} radiation-reaction forces acting upon the constituents of the binary system. Moreover, adding even more contrast to the electromagnetic case, the difficulty of dealing with these hereditary terms is further exacerbated by the (in)famous time nonlocality from tail effects, which introduces ultraviolet (UV) {\it and} infrared (IR) divergences in intermediate computations~\cite{Damour:2014jta,Galley:2015kus,Bernard:2015njp,Damour:2016abl,Marchand:2017pir,Foffa:2019rdf,Foffa:2019yfl}.~The divergences~are directly linked to the split into ``near" and ``far" zones, which are bread and butter of various perturbative expansions \cite{Porto:2017dgs}. The presence of IR poles, in particular, led to various discrepancies and {\it ambiguity parameters} in the original derivations of the 4PN conservative dynamics (e.g., see the discussion in \cite{Damour:2016abl}). These were ultimately resolved via a careful separation  between ``potential" and  ``radiation" regions within dimensional regularization (dim. reg.)~\cite{Porto:2017dgs}, amusingly similar to the Lamb shift \cite{Porto:2017shd}, yielding ambiguity-free results---entirely within the confines of the PN scheme---both in the traditional and EFT approaches \cite{Galley:2015kus,Marchand:2017pir,Foffa:2019rdf,Foffa:2019yfl}.\vskip 4pt
 
The hereditary story at the subsequent 5PN order had not concluded, until now, in a similar fashion. On the one hand, several contributions are well understood. Higher PN-order terms due to the leading (mass-type) tail \cite{Galley:2015kus}, as well as from higher order multipoles (e.g. the octopole \cite{Foffa:2019eeb}, etc.) are straightforward. Moreover, after a careful study of the multipole-moment decomposition in $d$ dimensions \cite{Henry:2021cek,Amalberti:2023ohj},  current-type tail terms were also derived to 5PN order \cite{Almeida:2021xwn,Blumlein:2021txe}, and agree in the overlap with the value inferred from the so-called {\it Tutti-Frutti} (TF) approach \cite{Bini:2019nra,Bini:2020wpo,Bini:2021gat}, and also with the recent 5PM (conservative) results at first order in the self-force expansion \cite{Driesse:2024xad}. On the other hand, the various values in the previous EFT literature in the PN regime \cite{Blumlein:2020pyo,Blumlein:2021txe,Almeida:2022jrv,Almeida:2023yia} for the failed tail (involving the total angular momentum) and  memory contribution (involving the product of three quadrupole moments) led to conflicting results, not only when compared against the TF approach \cite{Bini:2020wpo,Bini:2021gat} but also the total~\cite{Dlapa:2022lmu}  and conservative \cite{Dlapa:2021vgp,Bern:2021yeh} ${\cal O}(G^4)$ scattering angle obtained using EFT \cite{Kalin:2020mvi,Kalin:2022hph} and amplitude \cite{Bern:2019crd} methodologies. One of the main purposes of this paper is therefore to restore the harmony between the PN and PM derivations within the unifying EFT framework.\vskip 4pt 

A key element of our derivation is the role of diffeomorphism invariance. This demands that the multipole moments of the effective theory, which are defined in a locally-flat frame, are themselves subject to gravitational effects due to GW emission. Furthermore, although it turns out to be relevant at higher PN orders, the requirement of (manifest) gauge invariance of the long-distance theory forces upon us the use of the background-field gauge \cite{Goldberger:2004jt,Goldberger:2009qd}. These conditions guarantee that the Ward identities are automatically obeyed once the field equations for the complete two-body dynamics are enforced, and vice versa. Another important aspect of our computation is the extension of the EFT approach to the Schwinger-Keldysh ``in-in" formalism \cite{Keldysh:1964ud,Calzetta:1986cq}, which has already been proven to be very successful to incorporate dissipative effects in the PN and PM regimes, e.g. \cite{Galley:2009px,Galley:2010es,Galley:2012qs,Goldberger:2012kf,Galley:2013eba,Galley:2014wla,Galley:2015kus,Maia:2017yok,Maia:2017gxn,Goldberger:2020fot,Goldberger:2020wbx,Kalin:2022hph,Dlapa:2022lmu,Leibovich:2023xpg}. This requires a doubling of degrees of freedom, schematically $(x\to x_\pm, h \to h_\pm)$, for all of the worldline and bulk variables. Crucially, this is mandatory even for ``conserved" quantities, like the mass/energy, $M_\pm$, as well as the angular momentum, $L_\pm$, such that both must be included {\it and} varied through the Euler-Lagrange procedure. In addition, as we mentioned earlier, a well-known property of nonlinear gravitational interactions is the appearance of nonlocal-in-time effects, for instance, the well-known memory correction to the radiated angular momentum \cite{Arun:2009mc}. This feature must therefore also find its counterpart in the near-zone effective theory, and we demonstrate here the existence of nonlocal-in-time memory effects in the (in-in) effective action, captured by {\it boundary} contributions associated with {\it soft}-frequency limits of the Feynman integrals. After incorporating all of the aforementioned subtleties we complete the knowledge of hereditary effects in the two-body dynamics at 5PN order.\vskip 4pt 

As it was argued in \cite{Kalin:2020mvi}, a conservative-like contribution can be identified through the ``in-out" effective action, using Feynman's $i0^+$-prescription (while retaining the real part of the answer). 
For the case of tails and failed tails, the derivation of the conservative part is relatively straightforward, provided all the relevant contributions are included, and we agree with the results in~\cite{Almeida:2023yia,Henry:2023sdy}. On the other hand, we disagree with the conservative memory terms in \cite{Foffa:2019eeb,Blumlein:2021txe,Almeida:2023yia}, already at 4PM order. Furthermore, at 5PM and beyond, Feynman's prescription may introduce additional nonlocal-in-time effects, which were overlooked in previous derivations. The~existence of conservative-like hereditary corrections in gravity, however, is not the end of the story. There are other types of nonlinear contributions, namely those at second order in the leading radiation-reaction force.\footnote{Although the relevance of such terms was also pointed out in \cite{Bini:2021gat}, they had not been included until now in the derivation of conservative effects.} Moreover, their existence is implicit also for electromagnetic interactions, for which second-order effects in the  Abraham-(Dirac)-Lorentz force are responsible for conservative contributions in (relativistic) scattering computations~\cite{Bern:2023ccb}. We reproduce---from the point of view of the EFT in the PN scheme \cite{Goldberger:2009qd,Galley:2010es}---the leading order conservative-like radiation-reaction-{\it square} result reported in \cite{Bern:2023ccb}, and apply the same procedure to the gravitational ``Burke-Thorne" force \cite{Burke:1970dnm,Galley:2009px,Galley:2015kus}.\vskip 4pt 

After adding up all the relevant terms in the (in-in) effective action at 5PN order, including the known potential-only and tail-type corrections \cite{Blumlein:2020pyo,Almeida:2021xwn}, we derive the contribution to the total (even-in-velocity) relative scattering angle at ${\cal O}(G^4)$.  Perfect agreement is found in the overlap with the complete 4PM results~\cite{Dlapa:2022lmu}, as well as with the conservative part~\cite{Dlapa:2021vgp,Bern:2021yeh}. The computations reported here thus resolve the (apparent) discrepancy between the derivations in \cite{Dlapa:2022lmu,Dlapa:2021vgp,Bern:2021yeh,Bini:2020wpo,Bini:2021gat} and those in \cite{Foffa:2019eeb,Blumlein:2021txe,Almeida:2022jrv,Almeida:2023yia}, where the main differences between our present results and the latter can be traced to: {\it i)} The additional coupling between the (locally-flat) multipole moments and the gravitational field (yielding a different value for the ``double-bubble" diagram), {\it ii)} The inclusion of both $\pm$ contributions from ``conserved" quantities in the in-in action, ${\it iii)}$ The inclusion of nonlocal-in-time (boundary) terms, as well as {\it iv)} The proper identification of conservative effects through a `Principal Value' ($\PV$) prescription, and {\it v)} The inclusion of conservative-like radiation-reaction-square effects, all of which played a key role to achieve the aforementioned agreement at 5PN/4PM order. We will return to~the complete conservative sector at 5PN elsewhere.
The rest of this paper is organized as follows:\vskip 4pt 
 In~\S\ref{sec:ininEFT}, we briefly review the EFT approach and the in-in formalism. We emphasize the invariance under diffeomorphisms, and the need of a background-field gauge. In~\S\ref{sec:Tmn} we derive the contribution to the stress-energy tensor due to hereditary effects. We demonstrate the validity of the Ward identities for sources satisfying the expected GW fluxes at leading order. We then derive the energy and angular-momentum GW flux due to nonlinear gravitational effects. In~\S\ref{sec:rad} we derive the hereditary contributions to the (in-in) effective theory from failed-tail and memory effects.  We demonstrate the existence of nonlocal-in-time corrections arising as boundary terms in the soft-frequency limit. We also provide expressions for the nonlinear radiation-reaction forces, and explicitly show the equivalence between near and far-zone dissipative effects, including the known nonlocal-in-time contribution to the flux of (canonical) angular momentum. In~\S\ref{sec:cons} we introduce the conservative (in-out) effective action. We demonstrate the appearance, starting at 5PM order, of nonlocal-in-time effects due to Feynman's prescription, and identify the (local-in-time) contribution at~${\cal O}(G^4)$. In~\S\ref{sec:angle} we derive the impulse from all of the hereditary radiation-reaction forces, as well as all second order effects in the Burke-Thorne force, and their associated contribution to the total (relative) scattering angle at ${\cal O}(G^4)$. We find perfect consistency with the results first reported in~\cite{Dlapa:2022lmu} at 4PM order. We also discuss the conservative part, including failed-tail, memory, as well as radiation-reaction-square terms, finding as well agreement in the overlap with the value in~\cite{Dlapa:2021vgp,Bern:2021yeh}.  We conclude in~\S\ref{sec:disc} with a discussion on various subtleties in our derivations.~Other relevant aspects of our computations, including conservative-like radiation-reaction-square effects in electromagnetism and the role of the background-gauge fixing, are relegated to appendices. 

\section*{List of conventions}

\begin{itemize}
	\item We use the mostly minus signature $\eta_{\mu\nu} = \textrm{diag}(+, -, -, -)$ for the Minkowski metric.
	\item $\hbar = c = 1$, $ \kappa = \sqrt{32 \pi G} = m_{ \textrm{pl}}^{-1}$.
	\item  $\ddl{}^n(x) = (2\pi)^n\delta^n(x)$.
	\item We use Einstein's conventions for summations over repeated indices. To avoid confusion with the choice of metric signature, we use the Euclidian ${\bf 3}$-metric whenever results are written with space-like indices irrespectively of their (up or down) position.  
	\item We use (square) round brackets to identify a group of totally (anti-)simmetrized indices, e.g.
		\[
			A_{(\mu|} C_{\rho} B_{|\nu)} = \frac{1}{2}\big(A_\mu C_\rho B_\nu + A_\nu C_\rho B_\mu\big) \,,
			\qquad 
			A_{[\mu|} C_{\rho} B_{|\nu]} = \frac{1}{2}\big(A_\mu C_\rho B_\nu - A_\nu C_\rho B_\mu\big) \, .
		\]
	\item We work with dim. reg. in $d= 3-2\epsilon$ dimensions, and use the following shorthand for the $d+1$ and $d$ dimensional integrals,
		\[
		\int_{k, q, \cdots} \equiv \int \frac{\dd^{d+1} k}{(2\pi)^{d+1}}\frac{\dd^{d+1} q}{(2\pi)^{d+1}}\cdots \, ,
		\qquad
		\int_{\bk, \bq, \cdots} \equiv \int \frac{\dd^{d} \bk}{(2\pi)^{d}}\frac{\dd^{d} \bq}{(2\pi)^{d}}\cdots
		\]

	\item We use the convention
\begin{equation}
	I\ord{n}_{ab}(t) \equiv \frac{\dd ^n I_{ab}(t)}{\dd t^n}
\end{equation}
for the time derivatives of the quadrupole moment(s).
	\item We use the convention
	\beq
	f(x) = \int_k f(k) e^{-ik \cdot x}\,,
	\eeq
	for the Fourier transform.
	
\end{itemize}

\newpage
\section{The (in-in) EFT approach} \label{sec:ininEFT}

We briefly review the construction of the EFT approach and its extension to the Schwinger-Keldysh ``in-in" formalism below, see \cite{Porto:2016pyg,Goldberger:2022ebt} for further details.\vskip 4pt

The total effective action describing the two-body binary system interacting with long-wavelength gravitational fields takes the form
\begin{equation}
	S = S_{\textrm{EH}} + S_{\textrm{source}} \, ,
	\label{eq:Seff}
\end{equation}
where $S_{\textrm{EH}}$ is the standard Einstein-Hilbert term, 
\beq
S_{\rm EH} = -\frac{2}{\kappa^2} \int d^{d+1} x \sqrt{-g} R\,,
\eeq
and the source part given by \cite{Goldberger:2005cd,Goldberger:2009qd}
\begin{align}
		S_{\textrm{source}} & = - \int \dd \lambda \,  \bigg\{ \sqrt{g_{\mu\nu} V^\mu(\lambda) V^\nu(\lambda)} M(\lambda) + \frac{1}{2} \omega^{ab}_\mu L_{ab}(\lambda)V^\mu(\lambda)  - \frac{1}{2} I^{ab}(\lambda) \frac{E_{ab}}{\sqrt{g_{\mu\nu}V^\mu V^\nu} }+ \cdots \bigg\}\, ,
		\label{eq:StartingAction}
\end{align}
with $\big\{M(\lambda), L^{ab}(\lambda),I^{ab}(\lambda),\cdots \big\}\in \mathscr{M}^L$, the mass/energy, angular-momentum, symmetric-trace-free (STF) quadrupole moment, etc., of the binary system, including also binding (potential) degrees of freedom, and must be obtained through a matching computation \cite{Goldberger:2009qd}. The ellipses account for higher-order multipoles (as well as `finite-size' effects) which are not relevant for our purposes here. Greek indices, $\mu,\nu,\ldots$,  represent spacetime components, while the latin ones, $a,b,\ldots$, are local tensors projected through a tetrad field, $e^\mu_a$, with $e_0^\mu = V^\mu$, and obeying 
$g^{\mu\nu} = e_0^\mu e_0 ^\nu - \delta^{ab}e_a^\mu e_b^\nu \,$. The time variable $\lambda$ is any affine parameter for the dynamics of the center-of-mass worldline describing the binary system, $X^\mu(\lambda)$, with $V^\mu \equiv \dd X^\mu/ \dd\lambda$ its four-velocity. It is convenient to choose $\lambda \equiv X^0 = t$, and consider only the relative part of the full dynamics, which is captured by ignoring {\it recoil} effects, such that $X^\mu(t) = (t,0,0,0)$  and $V^\mu = (1,0,0,0)$. By performing a Lorentz transformation, we choose the tetrad to be nonrotating with respect to observers at infinity. The rotation of the binary is then described by the angular-momentum tensor, which couples to the gravitational field via the spin connection, $\omega^{ab}_\mu$, defined as usual,
\begin{equation}
	\omega^{ab}_\mu \equiv g^{\rho\sigma}e^b_\sigma  e^a_{\rho;\mu} \, .
\end{equation}
The quadrupole moment couples to the electric part of the Weyl tensor, $\mathcal{C}_{\mu\rho\nu\sigma}$, projected into the local frame,
\begin{align}
	E_{ab} \equiv e_a^\mu e^\nu_b \mathcal{C}_{\mu\rho\nu\sigma} V^{\rho} V^{\sigma} \, .
\end{align}
We choose the supplementary condition $L^{\mu\nu}V_\mu=0$ for the rotational degrees of freedom, which then translates into $L^{\mu\nu}e_{\mu}^0 = L^{\mu0}=0$ in the local frame. The same condition applies to the quadrupole moment, namely $I^{a0}=0$, since $E_{\mu\nu}V^\mu=0$. We introduce the angular-momentum vector, defined through $\bL^i = (1/2)\varepsilon^{i jk} L^{jk}$, contracted with the Euclidean ($\delta^{ij}$) metric, and $\varepsilon^{i jk}$ is the three-dimensional (flat) Levi-Civita symbol (with $\varepsilon^{123}=+1$).\vskip 4pt We split the metric into a background piece plus a perturbation, 
\begin{equation}
	g_{\mu\nu} = \bar{g}_{\mu\nu} + \kappa h_{\mu\nu} \, , \qquad \qquad 
	\bar{g}_{\mu\nu} =  \eta_{\mu\nu} + \kappa \bar{h}_{\mu\nu} \, .
\end{equation}
together with the tetrad,
\begin{align}
	e^\mu_a & = \eta_{a\rho}\bigg[\eta^{\rho\mu} 
		- \frac{\kappa}{2} \big(h^{\rho\mu} + \bar{h}^{\rho\mu}\big)
		+ \frac{3\kappa^2}{4} h^{\alpha(\rho} \bar{h}^{\nu)}{}_{\alpha} +\cdots  \bigg] \, , \\
	e_\mu^a & = \eta^{a\rho}\bigg[\eta_{\rho\mu} 
		+ \frac{\kappa}{2} \big(h_{\rho\mu} + \bar{h}_{\rho\mu}\big)
		- \frac{\kappa^2}{4} h^\alpha{}_{(\rho} \bar{h}_{\nu)\alpha} +\cdots \bigg] \, .
\end{align}
and use the following gauge-fixing term
\begin{equation}
	S_{\textrm{GF}} = \int \dd^{d+1} x \sqrt{-\bar{g}} \bar{g}^{\mu\nu}
		\left[\bar{g}^{\alpha\beta}\bar{\nabla}_{\alpha} h_{\beta\mu} - 
		\frac{\bar{g}^{\alpha\beta}}{2}\bar{\nabla}_\mu h_{\alpha\beta}\right]
		\left[\bar{g}^{\rho\sigma}\bar{\nabla}_{\rho} h_{\sigma\nu} - 
		\frac{\bar{g}^{\rho\sigma}}{2}\bar{\nabla}_\nu h_{\rho\sigma}\right] \, ,
		\label{eq:BGGF}
\end{equation}
with $\bar\nabla$ the covariant derivative, obeying $\bar\nabla_\sigma \bar g_{\mu\nu}=0$, which then preserves the gauge invariance under transformations of the background metric \cite{DeWitt:1967ub,tHooft:1974toh,Abbott:1980hw}.  As we shall see, the form of the last term in \eqref{eq:StartingAction} plays a crucial role enforcing diffeomorphism invariance, and ultimately the Ward identities, of the two-body system.\vskip 4pt

In order to incorporate dissipative effects, we implement the in-in formalism \cite{Keldysh:1964ud,Calzetta:1986cq}. This entails a doubling of the degrees of freedom, introducing a closed-time path action \cite{Galley:2009px,Galley:2012qs,Goldberger:2012kf,Galley:2013eba,Galley:2014wla,Galley:2015kus,Maia:2017yok,Maia:2017gxn,Kalin:2022hph,Dlapa:2022lmu,Leibovich:2023xpg},
\begin{equation}
S^{c} \equiv S_1 - S_2 = S_{\textrm{EH}, 1} + S_{\textrm{source}, 1} - S_{\textrm{EH}, 2} + S_{\textrm{source}, 2} \, .
\end{equation}
We will use Keldysh's parametrization by using the $\pm$ variables (for any field $\Phi$)  
\begin{equation}
	\Phi_+ = \frac{\Phi_1+\Phi_2}{2}  \, , \qquad \qquad \Phi_- = \Phi_1 - \Phi_2 \, .
\end{equation}
We compute the effective action, $\Gamma[\bar{h}_\pm, \mathscr{M}^L_\pm]$, by performing a path integral over the $h_{\pm, \mu\nu}$ field(s),
\begin{align}
	\exp\left\{i \Gamma[\bar{h}_\pm, \mathscr{M}^L_\pm]\right\} = \int \mathscr{D}[h_+,h_-] \exp\left\{i S^{c}[\bar{h}_\pm, h_\pm, \mathscr{M}^L_\pm] + S^{c}_{\textrm{GF}}[\bar{h}_\pm, h_\pm]\right\}  \, ,
\end{align}
which takes the form
\begin{align}
	\Gamma[\bar{h}_\pm, \mathscr{M}^L_\pm] = S_{\textrm{EH}}[\bar{h}_\pm] -\frac{\kappa}{2} \int \dd^{d+1} x \Big( \bar T_{+}^{\mu\nu} \bar{h}_{-,\mu\nu} + \bar T_{-}^{\mu\nu} \bar{h}_{+,\mu\nu} \Big) + O(\bar{h}_\pm^2) \, ,
\end{align}
with the stress-energy tensor given by,
\begin{equation}
	\bar T_{\pm}^{\mu\nu}  [\mathscr{M}^L_\pm]= -\frac{2}{\kappa}\frac{\delta \Gamma[\bar{h}_\pm, \mathscr{M}^L_\pm]}{\delta \bar{h}_{\mp, \mu\nu} (x)}
	 \bigg|_{\bar h_\pm\to 0} \, ,
	\label{eq:Tmunu}
\end{equation}
which is automatically conserved provided the sources satisfy the equations of motion. For the Feynman diagrams and rules we use the following conventions
\begin{align}
	 \raisebox{2pt}{\gravhb} =  h_\pm \, , & & 
	 \raisebox{2pt}{\gravh} = \bar{h}_\pm \, , & &
	 \raisebox{2pt}{ \Csoure} = \mathscr{M}_\pm^{ab\cdots}
\end{align}
with the standard retarded/advanced propagators,%
\begin{align}
	\retg = \frac{i}{(k^0 + i0^+)^2 - \bk^2} P_{\mu\nu\rho\sigma} \, ,  & & 
	\advg = \frac{i}{(k^0 - i0^+)^2 - \bk^2} P_{\mu\nu\rho\sigma} \, ,  
\end{align}
and $P_{\mu\nu\rho\sigma} 
\equiv \eta_{\mu(\rho}\eta_{\sigma)\nu} - \tfrac{1}{d-1}\eta_{\mu\nu}\eta_{\rho\sigma}$. For our purposes, we just require the cubic-vertex interaction
\begin{align}
	\cubicbackpp & = i \kappa \int_k \ddl{}^{d+1}(k_1+k_2-k) V_{++}^{\alpha_1 \beta_1 \alpha_2\beta_2}{}_{\mu\nu} h^{\mu\nu}_-(k) \, , 
\end{align}
and for the sources,
\begin{align}
	\Mphb \, , & &  \Mph \, , & &  \Mmh \, ,\\
	\Mphbtwo \, , & &  \Mmhtwo \,  & & ,
\end{align}
where, to the order we work in this paper, we have (in $d=3$)
\beq
\begin{aligned}	
M_\pm &= \sum_a m_a\,,\\
L_-^{ij} &= 2\sum_a m_a\bigg ( \bx_{a,-}^{[i}\bv_{a,+}^{j]} + \bx_{a,+}^{[i}\bv_{a,-}^{j]} \bigg)\,,\\
L_+^{ij} &= 2\sum_a m_a \bx_{a,+}^{[i}\bv_{a,+}^{j]}\,,\\
I_-^{ij} &= \sum_a m_a \bigg( 2 \bx_{a, -}^{(i}\bx_{a, +}^{j)}- \frac{2}{3}\delta^{ij}\bx_{a, -}\cdot \bx_{a, +} \bigg) \,,\\
I_+^{ij} &= \sum_a m_a \bigg(  \bx_{a, +}^{i}\bx_{a, +}^{j}- \frac{1}{3}\delta^{ij}\bx_{a, +}\cdot \bx_{a, +} \bigg)\,,
\end{aligned}
\eeq
for the relevant multipole moments. In what follows we concentrate on the nonlinear corrections involving the angular-momentum and quadrupole couplings.

\section{Stress-energy tensor} \label{sec:Tmn}

\begin{figure}[t]
\begin{center}
	\includegraphics{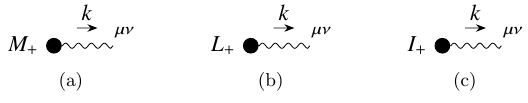}
\end{center}
\caption{The  Feynman diagrams needed for the computation $T_+^{\mu\nu}$ at leading order.}
\label{Fig:LOT}
\end{figure}

For the derivation of the stress-energy tensor we consider contributions to $\bar T_{+}^{\mu\nu}$, which is the only relevant component in the classical limit (in which we ignore closed loops of the gravitational field). We compute all (tree-level) connected Feynman diagrams with one external $\bar{h}_{-, \mu\nu}$. At leading order we have the diagrams in Fig.~\ref{Fig:LOT}, and we get in momentum space 
\beq
\begin{aligned}
	\bar T_{+\rm (LO)}^{\mu\nu}(k) = &\left( M_+(\omega)-\frac{1}{2}I_+^{ab}(\omega)\delta^{\rho}_a \delta^{\sigma}_b k_\rho k_\sigma \right)V^\mu V^\nu + \\  &\delta^a_\nu\delta^b_\rho V^{(\mu}  \left( \omega I_+^{a)b}(\omega) + i L_+^{a)b}(\omega)  \right)k_\rho  V^{\nu)}-\delta^{\mu}_a\delta^{\nu}_b \frac{\omega^2}{2}I_+^{ab}(\omega) \,.
	\label{eq:TpLO}
\end{aligned}
\eeq
In order to avoid cluttering of notation, in what follows we will use an abuse of notation and sometime also utilize Greek letters for the indices of the $\mathscr{M}^L$'s. The reader should keep in mind that these variables are ultimately projected onto the local frame using the Euclidean ${\bf 3}$-metric. We will also remove the bar on $\bar T^{\mu\nu}$.  We discuss next the correction due to hereditary effects.

\subsection{Failed tail \& Memories}

\begin{figure}[t]
\begin{center}
	\includegraphics{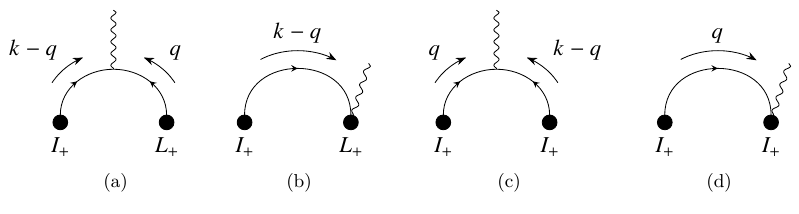}
\end{center}
\caption{The Feynman diagrams needed for the NLO $T_+^{\mu\nu}$ in the physical limit. Diagrams (a) and (b) give the failed tail contribution, while (c) and (d) are responsible for the memory term. (We did not include the symmetric version of diagram (b), since it vanishes.)}
\label{Fig:NLOT}
\end{figure}

The failed-tail contribution is given by the diagrams (a) and (b) in Fig.~\ref{Fig:NLOT}. The derivation entails an integral of the form 
\begin{equation}
	T^{\mu\nu}_{+, (\textrm{FT})}(k) = \kappa^2 L_+^{\alpha\beta} I_+^{\rho\sigma}(\omega)\int_q 
		\frac{\ddl(q^0) \mathcal{N}\ord{\rm FT}_{\alpha\beta\rho\sigma}{}^{\mu\nu}(k, q, v, \eta)}{\bq^2 [ (\omega +i0^+)^2 -|\bk - \bq|^2 ]} \, ,	
		\label{eq:Ftailschem}
\end{equation}
where we have already used that  $L_+^{\alpha\beta}(q^0) = L_+^{\alpha\beta} \ddl(q^0) + {\cal O}(G^2)$. Using tensorial reduction and integration-by-parts (IBP) identities, we get
\begin{equation}
T^{\mu\nu}_{+ \textrm{(FT)}}(k) = \kappa^2 L_+^{\alpha\beta} I_+^{\rho\sigma}(\omega) \Big[ \mathcal{N}\ord{{\rm FT}_1}_{\alpha\beta\rho\sigma}{}^{\mu\nu}(k, v, \eta) {\cal I}^{(\rm FT)}_{0, 1} + \mathcal{N}\ord{{\rm FT}_2}_{\alpha\beta\rho\sigma}{}^{\mu\nu}(k, v, \eta) {\cal I}^{(\rm FT)}_{1, 1} \Big] \, .
\label{eq:failtailschem}
\end{equation}
where we introduced the following family of master integrals 
\begin{equation}
	{\cal I}^{\rm (FT)}_{ab} \equiv \int_{\bq} \frac{1}{[\bq^2]^a [ (\omega +i0^+)^2 -|\bk - \bq|^2 ]^b} \, .
\end{equation}
\vskip 4pt 

Next we move to the computation of the memory contribution. This is obtained by computing the diagrams (c) and (d) of Fig.~\ref{Fig:NLOT}. Schematically, we find
\begin{equation}
	T^{\mu\nu}_{+, (\textrm{M})}(k) = \kappa^2 \int_q 
		\frac{\mathcal{N}\ord{\rm M}_{\alpha\beta\rho\sigma}{}^{\mu\nu}(k, q, v, \eta)}{\big[ (q^0 +i0^+)^2 -|\bk|^2 \big] \big[ (\omega - q^0 +i0^+)^2 -|\bk - \bq|^2 \big]} I_+^{\alpha\beta}(q^0) I_+^{\rho\sigma}(\omega - q^0) \, .	
\end{equation}
Through the use of IBP and tensorial reduction, this result can be simplified to 
\begin{align}
T^{\mu\nu}_{+ \textrm{(M)}}(k) = \kappa^2  \int \frac{\dd q^0}{2\pi} I_+^{\rho\sigma}(q^0) I_+^{\alpha\beta}(\omega - q^0)  \Big[ & \mathcal{N}\ord{{\rm M}_1}_{\alpha\beta\rho\sigma}{}^{\mu\nu}(k, q^0, v, \eta) \mathcal{I}^{(\rm M)}_{1, 0} 
	+ \mathcal{N}\ord{{\rm M}_2}_{\alpha\beta\rho\sigma}{}^{\mu\nu}(k, q^0, v, \eta) \mathcal{I}^{(\rm M)}_{0, 1} \notag \\
	& 
	+ \mathcal{N}\ord{{\rm M}_3}_{\alpha\beta\rho\sigma}{}^{\mu\nu}(k, q^0, v, \eta) \mathcal{I}^{(\rm M)}_{1, 1} \Big] \, ,
	\label{eq:Tmemschem}
\end{align}
where, likewise, we have introduced the following family of master integrals
\begin{equation}
	\mathcal{I}^{(\rm M)}_{ab} \equiv \int_{\bq} \frac{1}{\big[ (q^0 +i0^+)^2 -|\bk|^2 \big]^a \big[ (\omega - q^0 +i0^+)^2 -|\bk - \bq|^2 \big]^b} \, .
\end{equation}
The explicit form of the stress-energy tensor, after adding up both contributions, is not particularly illuminating. We will use it shortly to derive the far-zone metric. Nevertheless, it is instructive to check the conservation laws, which we do next. 

\subsection{Ward identity} 

We now verify that the stress-energy tensor computed previously satisfies the Ward identities, 	$\partial_\mu T_+^{\mu\nu}(x) = 0$,
up to the order we are interested; or equivalently, in Fourier space, 
\begin{equation}
	\partial_\mu T_+^{\mu\nu}(x) = - i \int_k k_\mu T_+^{\mu\nu}(k) e^{-i k\cdot x} = 0 \, .
\end{equation}
From the result in \eqref{eq:TpLO}, we find %
\begin{equation}
	\partial_\mu T^{\mu\nu}_{+, {(\rm LO)}}(x) = \dot{M}_+(t) \delta^3(\bx) + \frac{1}{2}\dot{L}_+^{\mu \alpha}(t) \partial_\alpha \delta^3(\bx) = 0 +{\cal O}(G^2) \, ,
	\label{eq:consLO}
\end{equation}
 at leading order in $G$, as expected.\vskip 4pt Moving onto hereditary effects, for the failed tail in \eqref{eq:failtailschem} we readily find that $k_\nu T^{\mu\nu}_{+ \textrm{(FT)}}(k)$ vanishes. Notice that, while the seagull-type (double-bubble) diagram in Fig.\ref{Fig:NLOT}(b) is not present in tail-type terms, it is crucial for the contribution from the failed tail to guarantee the conservation law.\vskip 4pt 
 
 We are then left with the remaining combination
\begin{equation}
k_\nu \Big( T^{\mu\nu}_{+({\rm LO})}(k) + T^{\mu\nu}_{+ \textrm{(M)}}(k) \Big) \, ,
\end{equation}
which includes the terms from~\eqref{eq:TpLO} and \eqref{eq:Tmemschem}, evaluated at the next order in $G$. After contracting with $k_\mu$, the result for the memory part greatly simplifies, using 
\beq
{\cal I}^{\rm (M)}_{1,0} = - \frac{i}{4\pi} q_0 + {\cal O}(d-3) \, , \qquad {\cal I}^{\rm (M)}_{0,1} = - \frac{i}{4\pi} (w-q_0) + {\cal O}(d-3) \,,
\eeq
for the relevant integrals, we find
\begin{align}
	k_\nu T^{\mu\nu}_{+ \textrm{(M)}}(k) = - i \frac{G}{5} \int \frac{\dd q^0}{2\pi}  \bigg\{ &  q_0^5 (\omega - q_0) I^{\alpha\beta} (q_0)I_{\alpha\beta}(\omega - q_0) v^\mu \notag \\
	& - \Big[\big(\omega - q_0\big)^5 - q_0^5  \Big]I^{\mu\alpha} (q_0)I_{\alpha\beta}(\omega -q_0) k^\beta    \bigg \} \, ,
	\label{eq:FTcons}
\end{align}
where the first line only contributes to the $k_\nu T_+^{\nu 0}$ component, whereas the second goes into the spatial part. In~coordinate space, this becomes 
\begin{align}
	\partial_\nu T^{\nu\mu}_{+ \textrm{(M)}}(x) & =  \frac{G}{5} \bigg( I\ord{5}_{ij}(t)I\ord{1}_{ij}(t) \delta^\mu_0 \bigg) \delta^3(\bx) + \frac{G \delta^{\mu}_i }{5} \bigg({I}^{i}{}_{j}(t)I\ord{5}_{jk}(t)-{I\ord{5}}^{i j}(t)I_{jk}(t)\bigg) \partial_k\delta^3(\bx) \,,
\end{align}
 Hence, including the leading order terms coming from \eqref{eq:consLO}, we arrive at 
\beq
\begin{aligned}
	\partial_\nu T_+^{\nu0}(x) & = \left(\dot{M}(t) + \frac{G}{5} I\ord{5}_{ij}(t)I\ord{1}_{ij}(t)\right) \delta^3(\bx) \, , \\
	\partial_\nu T_+^{\nu}{}_i(x) & = \left[\frac{1}{2}\dot{L}_{ik}(t) + \frac{G}{5} \bigg( I_{i j}(t)I\ord{5}_{jk}(t)-I\ord{5}_{i j}(t)I_{jk}(t)  \bigg)\right]\partial_k\delta^3(\bx) \,,
\end{aligned}
\eeq
to the desired order. The above expression vanishes upon using the (near-zone) conservation laws for the sources,\footnote{In principle we can also add the coupling to the total linear momentum in the effective action, see e.g.~\cite{Porto:2016pyg}, which would enter in the $0i$ component of the Ward identity, yielding the expected flux of radiated linear momentum proportional to the coupling between the octupole and the quadrupole moments. We can then add the associated radiation-reaction force, closing the ``self-energy" diagram, which would then account for the recoil effects we are not including here.} 
\begin{align}
	\dot{M}(t)  = -\frac{G}{5}I\ord{5}_{ij}(t)I\ord{1}_{ij}(t) \, , \quad
		
	\dot{L}_{ij}(t)  = -\frac{4G}{5} I_{k[i}(t)I\ord{5}_{j]k}(t)  \, ,
		\label{eq:consmem}
\end{align}
that follow from the leading order (in-in) effective action \cite{Galley:2015kus}. The energy conservation then agrees, as expected, with the result in~\cite{Goldberger:2012kf}, while the angular-momentum part extends it to the other components. Upon time averaging (i.e. up to {\it Schott} terms), we  reproduce the known values (see, e.g., Eqs. (3.75) and (3.97) of \cite{Maggiore:2007ulw})
\begin{align}
	\langle \dot{M}(t) \rangle = -\frac{G}{5}\left\langle I^{ij(3)}(t)I^{ij(3)}(t)\right\rangle \, ,\quad 
	\langle \dot{\bL}^{i}(t) \rangle = -\frac{2G}{5} \left\langle \varepsilon^{ijk} I^{lj(2)}(t)I^{lk(3)}(t) \right\rangle\, .
	\label{eq:conslaw}
\end{align}
Let us point out that, although the additional terms from the background-gauge condition did not feature in the Ward identity at this order, as we demonstrate in App.~\ref{app:WI}, the (covariant) gauge fixing plays an important role guaranteeing the conservation laws at higher orders.

\subsection{GW fluxes}
\label{sec:fluxeshmunu}

We compute now the radiated energy and angular momentum due to hereditary effects.  We start by deriving the asymptotic waveform in transverse-traceless (TT) gauge,  
\begin{equation}
	h^{\textrm{TT}}_{ij}(x_{\rm ret}) = \frac{\kappa}{8 \pi r} \int\frac{\dd\omega}{2\pi} 
		e^{-i \omega t_{\rm ret}} f_{ij}(\omega, \omega \bn) \, , 
	\qquad
	f_{ij}(\omega, \omega \bn) \equiv \Lambda_{ij ab} T_{ab, +}(\omega, \omega \bn) \, ,
	\label{eq:httOS}
\end{equation}
where $x_{\rm ret} \equiv (t_{\rm ret},\bx)$, evaluated on the retarded time, and we introduced the normalized radial direction $\bn$, with $\bn\cdot\bn = 1$, and $\Lambda_{ij ab}$ is given by
\begin{equation}
	\Lambda_{ij ab} = \Pp_{ia}\Pp_{jb} - \frac{1}{2}\Pp_{ij}\Pp_{ab} \, , \qquad \Pp_{ij} \equiv \delta_{ij} - n_i n_j \,,
\end{equation}
which serves as a projector onto the TT gauge gauge. In what follows we drop the `ret' label. From the asymptotic waveform, we compute the loss of energy and angular momentum (notice the overall minus signs)
\begin{align}
	\dot{M}(t) & = - \lim_{r \to \infty} \int \dd \Omega \, r^2 
		 \dot{h}^{\textrm{TT}}_{ij} \dot{h}^{\textrm{TT}}_{ij} \, ,
		\label{eq:MdotTT}\\ 
	\dot{L}_{ij}(t) & = - \lim_{r \to \infty} \int \dd \Omega \, r^2 \left[ 
		2 h^{\textrm{TT}}_{a [i}\dot{h}^{\textrm{TT}}_{j] a} 
		- \dot{h}^{\textrm{TT}}_{ab} x_{[i}\partial_{j]} h^{\textrm{TT}}_{ab} 
	\right] \, .
\end{align}
The waveform receives contributions at leading order, $f_{(0)}^{ij}(\omega)$, and from hereditary effects, $ f_{\rm (H)}^{ij}(\omega)$, such that, for the energy flux we have  %
\begin{align}
	\dot{M}_{\rm (H)}(t) & = \frac{\kappa ^2}{32 \pi^2 }\int \dd \Omega \int \frac{\dd \omega_1 \dd \omega_2}{(2\pi)^2} e^{-i (\omega_1 + \omega_2) t} \omega_1\omega_2
	\Big[ f_{(0)}^{ij}(\omega_1) f_{\rm (H)}^{ij}(\omega_2)
		\Big]  \, .
		\label{eq:MdotNLO}
\end{align}
whereas for the angular-momentum flux, %
\begin{align}
	\dot{L}^{ij}_{\rm (H)}(t) & = \frac{i \kappa ^2}{64 \pi^2 }\int \dd \Omega \frac{\dd \omega_1 \dd \omega_2}{(2\pi)^2} e^{-i (\omega_1 + \omega_2) t} \big( \omega_2 -  \omega_1 \big)
	\Big[ 
		2f_{(0)}^{l [i}(\omega_1) f_{\rm (H)}^{j] l}(\omega_2) 
		- f_{\rm (H)}^{kl}(\omega_2) n^{[i}\tilde{\partial}^{j]} f_{(0)}^{kl}(\omega_1)
		\Big] \, ,
		\label{eq:LdotNLO}
\end{align}
where $x_{[i}\partial_{j]} = n_{[i}\Pp_{j]a}\tilde{\partial}{_a}$, with $\tilde{\partial}_a \equiv \partial/(\partial n^a)$. Using the result given in~\eqref{eq:Ftailschem}, we then get the following contribution to the energy loss due to the failed-tail coupling,%
\begin{align}
	\dot{M}_{\textrm{(FT)}}(t) & = -\frac{2 G^2}{15} L^{j k}\int \frac{\dd \omega_1 \dd \omega_2}{(2\pi)^2} e^{-i (\omega_1 + \omega_2) t} 
		I^{lj}(\omega_1)I^{lk}(\omega_2) \omega_1^3 \omega_2^5\notag\\
	& = -\frac{2 G^2}{15} L^{j k} I^{(3)lj}(t) I^{(5)lk}(t) = 0 + \frac{\dd}{\dd t}\big(\cdots \big)  \, .
	\label{eq:MdotflFT}
\end{align}
which vanishes at this order. We omit the expression of the total derivative, which cancels out upon time averaging. For the memory part we find  (with $\omega_3 = \omega_2-q_0$)
\beq
\begin{aligned}
	\dot{M}_{\textrm{(M)}}(t) & = \frac{2 iG^2}{35} \int \frac{\dd \omega_1 \dd \omega_2 \dd q_0}{(2\pi)^3} e^{-i (\omega_1 + \omega_2) t}		I^{ij}(\omega_1)I^{jk}(q_0)I^{ki}(\omega_3) \omega_1^3 q_0^3 
		  \big( 2 \omega_3^3 + 11 \omega_3^2 q_0 + 14 \omega_3 q_0^2 + 7 q_0^3\big)  \\
		& = \frac{G^2}{5}  I^{(1)ij}(t)I^{(4)jk}(t)I^{(4)ki}(t) + \frac{\dd}{\dd t}\big(\cdots \big)  \, .
		\label{eq:Mdotmemory}
\end{aligned}
\eeq
Following similar steps, the failed-tail contribution to the flux of angular momentum becomes
\begin{align}
	\dot{L}^{ij}_{\textrm{(FT)}}(t) & = \frac{2 G^2}{15} i\int \frac{\dd \omega_1 \dd \omega_2}{(2\pi)^2} e^{-i (\omega_1 + \omega_2) t} 
		\Big( I^{ki}(\omega_1)I^{jl}(\omega_2) L^{kl} + I^{kl}(\omega_2)I^{k[i}(\omega_1)  L^{j]l} \Big) \omega_1^2(\omega_1 -\omega_2) \omega_2^4 \notag \\
	& = \frac{4 G^2}{15} I^{(4)kl}(t)I^{(3)k[i}(t) L^{j] l} + \frac{\dd}{\dd t}\big( \cdots \big) \,.
	\label{eq:LdotFT}
\end{align}
There is an important difference for the memory part, which contains a term of the form
\begin{align}
	\dot{L}_{\rm nloc (M)}^{ij}(t) & = \frac{8G^2}{35}\int \frac{\dd \omega_1 \dd \omega_2 \dd \omega_3}{(2\pi)^3} e^{-i (\omega_1 + \omega_2) t}
		I^{k[i}(\omega_1)I^{j]l}(q_0)I^{kl}(\omega_3) \bigg[\frac{\omega_1^3 q_0^6}{\omega_2+i0^+} \bigg] \, ,
\end{align}
responsible for nonlocal-in-time effects in the flux. This can be seen by rewriting it as%
\begin{align}
	\dot{L}_{{\rm nloc (M)}}^{ij}(t) = \frac{8 G^2}{35} & 
	\int \dd \tau I^{(3)k[i}(t)
			I^{(6)j]b}(\tau) I^{kl}(\tau)\int \frac{\dd \omega_2}{2\pi} \frac{i e^{i \omega_2 (\tau- t)}}{\omega_2+i0^+} \, , 
\end{align}
and using %
\begin{equation}
	i \int \frac{\dd \omega_2}{2\pi} \frac{e^{i \omega_2 (t_3- t)}}{\omega_2 + i 0^+} = \vartheta(t - t_3) \, ,
\end{equation}
such that
\begin{align}
	\dot{L}_{\rm nloc (M)}^{ij}(t) = \frac{8 G^2}{35} I^{(3)k[i}(t) \int_{-\infty}^t \dd \tau \, 
			I^{(6)j]l}(\tau) I^{kl}(\tau) \,,
			\end{align}
which agrees with the known nonlocal-in-time contribution (see e.g. Eq.~(2.8) in \cite{Bini:2021qvf}).  Combining the pieces, we find 
\begin{align}
	\dot{L}^{ij}_{\rm(M)}(t) = & -\frac{4 G^2}{5} I^{(4)a[i}(t) I^{j] b}(t)I^{(4)ab}(t) -\frac{8 G^2}{35} I^{(2)a[i}(t) I^{(3)j] b}(t)I^{(3)ab}(t) \notag \\
		  & - \frac{8 G^2}{35} I^{(3)a[i}(t) \int_{-\infty}^t \dd \tau \, 
			I^{(3)j]b}(\tau) I^{(3)ab}(\tau) + \frac{\dd}{\dd t}\big( \cdots \big) \, , 
		\label{eq:LdotFluxmem}
\end{align}
where the last total derivative involves only local-in-time terms which vanish at infinity.\vskip 4pt For the sake of comparison with the literature,~e.g.~\cite{Arun:2009mc}, it is instructive to compute an averaged value by integrating over the binary's history divided by $T$, the elapsed time, and take the $T \to \infty$ limit. For the nonlocal-in-time part we find
\beq
\begin{aligned}
 \langle \dot{L}^{ij}_{\rm nloc(M)}(t)\rangle &\equiv \lim_{T \to \infty} \frac{1}{T}\int \dd t \dot{L}^{ij}_{\rm nloc(M)}(t) \\ &=  \frac{8G^2}{35 T} \int \frac{\dd \omega_1 \dd \omega_2 \dd q_0}{(2\pi)^3} \ddl(\omega_1 +\omega_2)
		 \bigg[
		\frac{\omega_1^3 q_0^6}{\omega_2 + i0^+}  \bigg]I^{a[i}(\omega_1)I^{j]b}(q_0)I^{ab}(\omega_2-q_0) \, .
		\label{eq:avLdotstep}
\end{aligned}
\eeq
Naively, one would be tempted to solve the $\delta(\omega_1+\omega_2)$ and simply cancel a factor of $\tfrac{\omega_1}{\omega_2+i0^+}$. However, that would be incorrect since that ignores the $\omega_2 \to 0$ soft-frequency limit, for which the $i0^+$-prescription becomes important. Using the distribution identity
\begin{equation}
	\frac{1}{\omega_2 + i0^+} = \frac{\PV}{\omega_2}- \frac{i}{2}\ddl(\omega_2) \, ,
	\label{eq:PVplusdelta}
\end{equation}
then~\eqref{eq:avLdotstep} can be split into two terms \beq \langle \dot{L}^{ij}_{\rm nloc(M)}(t)\rangle=  \langle \dot{L}^{ij}_{\rm nloc(M)}(t)\rangle_{\PV} + \langle \dot{L}^{ij}_{\rm nloc(M)}(t)\rangle_{\delta}.\eeq 
One the one hand, using the distributional identity  $\omega \PV(1/\omega) = 1$, the part of the principal value renormalizes the local-in-time contribution adding up to the total value
\begin{align}
	\langle \dot{L}^{ij}_{\rm loc(M)}(t)\rangle &= 
	-\frac{i}{T} \frac{4G^2}{35} \int \frac{\dd \omega_1 \dd \omega_2 \dd q_0}{(2\pi)^3}
		I^{a[i}(\omega_1)I^{j]b}(q_0)I^{ab}(\omega_2-q_0) \omega_1^3 q_0^6 \ddl(\omega_1+\omega_2)\ddl(\omega_2) \, \nn \\
&= -\frac{4 G^2}{5}\left\langle
		   I^{(4)a[i}(t)I^{j] b}(t)I^{(4)ab}(t) 
		  \right\rangle \, .
		  \label{eq:dotLML}
\end{align}
On the other hand, the nonlocal-in-time part involving the zero-frequency limit yields
\begin{align}
	\langle\dot{L}^{ij}_{\textrm{nloc(M)}}(t)\rangle_\delta & = 
	-i\frac{4G^2}{35T} 
	\int \frac{\dd \omega_1}{2\pi} \ddl(\omega_1) \omega_1^3I^{a[i}(\omega_1) 
	\int \frac{\dd q_0}{2\pi}
		I^{j]b}(q_0)I^{ab}(-q_0)  q_0^6  \notag \\
		& = -\frac{4 G^2}{35} \left \langle I^{(3)a[i}(t)\right\rangle 
		\int \dd \tau I^{(3)j]b}(\tau)I^{(3)ab}(\tau) \, .
		\label{eq:dotLMNL}
\end{align}
This results then agrees with the value in  \cite{Arun:2009mc}, where the nonlocal-in-time term is associated with a so-called `DC' memory contribution, see e.g. the first part of Eq. (5.14) in \cite{Arun:2009mc}, and notice that the factor of $1/2$ in~\eqref{eq:PVplusdelta} accounts for the {\it half} integration over the energy flux.

\section{Radiative action} \label{sec:rad}

In order to obtain the form of the radiation-reaction forces upon the binary's dynamics, we compute the total (in-in) effective action by integrating over the $
\bar h_\pm$ field. We perform the computation for the failed-tail and memory contributions in what follows.%
\begin{figure}[t]
\begin{center}
	\includegraphics[scale=0.9]{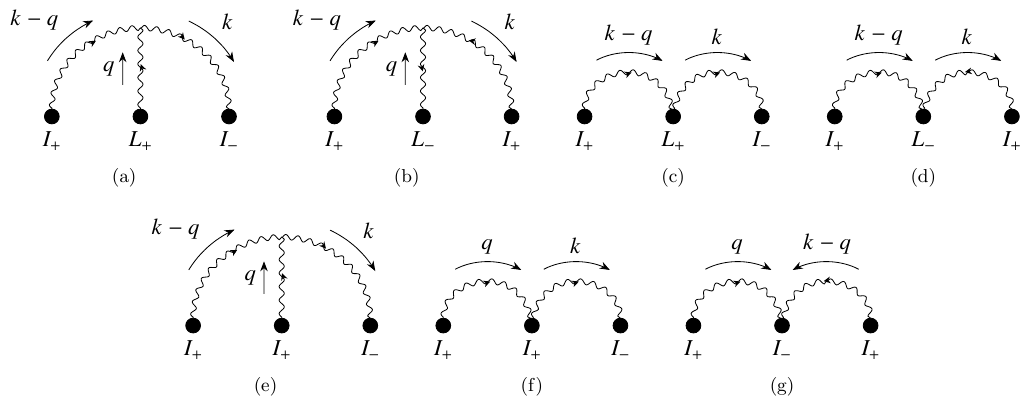}
\end{center}
\caption{Feynman diagrams contributing to the in-in effective action. Diagrams (a) to (d) enter in the failed tail, whereas the memory is given by the sum of diagrams (e), (f) and (g).}
\label{Fig:Sinin}
\end{figure}

\subsection{Failed tail}

The failed tail is given by the diagrams in Figs.~\ref{Fig:Sinin}(a) to \ref{Fig:Sinin}(d). After various manipulations, including  tensorial reduction and IBP relations, we find%
\begin{align}
	S_{(\textrm{FT})}^{\textrm{(a)}\&\textrm{(c)}} & = +\frac{1}{15}\frac{i\kappa^4}{64}\int \frac{\dd\omega\dd q_0}{(2\pi)^2} L_+^{kl}(q_0) I_-^{ik}(-\omega)I_+^{li}(\omega -q_0) \omega^5 \mathcal{I}_{(\textrm{FT})+}(\omega) \, , \\
	S_{(\textrm{FT})}^{\textrm{(b)}\&\textrm{(d)}} & = -\frac{1}{15}\frac{i\kappa^4}{128}\int \frac{\dd\omega\dd q_0}{(2\pi)^2}  L_-^{kl}(q_0)I_+^{ik}(\omega -q_0)I_+^{li}(-\omega) \omega^5 \mathcal{I}_{(\textrm{FT})-}(\omega) \, , 
\end{align}
where it is understood that, at this order, $L_+^{kl}(q_0) = L_+^{kl}\ddl(q_0)$. For this reason, the final result is really just a function of one frequency $\omega$. The two relevant master integrals, $\mathcal{I}_{(\textrm{FT})\pm}(\omega)$, are straightforward to compute,
\begin{equation}
	\mathcal{I}_{(\textrm{FT})\pm}(\omega) = \int_{\bk} \frac{1}{(\omega \pm i0^+)^2-\bk^2}\int_{\bq}\frac{1}{(\omega+i0^+)^2-\bq^2} = \mp \frac{\omega^2}{16\pi^2} + O(d-3) \, .
\end{equation}
After Fourier transforming to coordinate space, we have %
\begin{align}
	S_{(\textrm{FT})} & = -\frac{G^2}{15}\int \! \dd t \, L_{+}^{kl}I\ord{4}_{-,kj}I\ord{3}_{+, jl}
		 + \frac{G^2}{30}\int \! \dd t \, L_{-}^{kl} I\ord{4}_{+,kj}I\ord{3}_{+, jl}  \, .
	\label{eq:Sfail}
\end{align}
Notice that, unlike tail terms, e.g. \cite{Galley:2015kus}, the failed-tail contribution is finite in $d=3$.\vskip 4pt  The first term in \eqref{eq:Sfail} agrees with the result in \cite{Almeida:2023yia}. On the other hand, the term proportional to $L^{ij}_-$ was not included. As we shall see in \S\ref{sec:Flux}, this extra term is  crucial to recover the flux of angular momentum derived in~\eqref{eq:LdotFT}.

\subsection{Memories} 

The memory contribution is given by adding Figs.~\ref{Fig:Sinin}(e),~\ref{Fig:Sinin}(f) and~\ref{Fig:Sinin}(g). While the computation of the last two diagrams is straightforward, the Feynman integral in~\ref{Fig:Sinin}(e) turns out to be subtle, depending on the following family of master integrals
\begin{equation}
	{\cal I}_{abc} \equiv \int_{\bq, \bk} \frac{1}{[(\omega_1 +i0^+)^2-\bq^2]^a [(\omega_2 +i0^+)^2-|\bk +\bq|^2]^b [(\omega_3 +i0^+)^2-\bk^2]^c} \, ,
\end{equation}
such that
\begin{align}
	 S_{\rm (M)}^{\rm (e)} = \kappa^4 & \int \frac{\dd \omega_1 \dd \omega_2 \dd \omega_3}{(2\pi)^3} \ddl(\omega_1 + \omega_2 - \omega_3)
	I_-^{ij}(-\omega_3)I_+^{jk}(\omega_1)I_+^{ki}(\omega_2)  \bigg[ 
	c_1 {\cal I}_{110} + c_2 {\cal I}_{101} + c_3 {\cal I}_{011} +  c_4 {\cal I}_{111}
	\bigg] \, ,
	\label{eq:SNLstep}
\end{align}
where the $c_i$'s are functions of $\omega_{1,2,3}$. While the other diagrams in Figs.~\ref{Fig:Sinin}(f) and~\ref{Fig:Sinin}(g) only depend on the ${\cal I}_{110}$, ${\cal I}_{101}$ and ${\cal I}_{011}$ masters, which are straightforward to compute, i.e.
\begin{align}
	{\cal I}_{110} & =-\frac{\omega_1 \omega_2}{16\pi^2} +{\cal O}(d-3) \, , \,\, {\cal I}_{101}  = -\frac{\omega_1 \omega_3}{16\pi^2} +{\cal O}(d-3) \,, \,\, {\cal I}_{011} =-\frac{\omega_2 \omega_3}{16\pi^2} +{\cal O}(d-3) \,,\label{eq:011}
\end{align}
the expression in \eqref{eq:SNLstep} depends also on ${\cal I}_{111}$, which a priori entails three different frequencies. Naively, due to the overall $\delta$-function, one would be tempted to replace $\omega_1+\omega_2-\omega_3=0$, in which case ${\cal I}_{111}$ reduces through IBP relations into a combination of the ones in \eqref{eq:011}. Furthermore, although divergent ($\propto 1/(d-3)$), the ${\cal I}_{111}$ enters through a term proportional to $(\omega_1+\omega_2-\omega_3)^2$, dropping out of the final (finite) answer. Incidentally, this was also the strategy adopted in the derivations in \cite{Foffa:2019eeb,Blumlein:2021txe,Almeida:2022jrv}. As we shall see, once the corrected Feynman rules are utilized, this procedure accurately captures the local-in-time part (see below). However, it does not fully account for all the relevant terms. In particular, it fails to reproduce the nonlocal-in-time flux of angular momentum obtained from the one-point function. That is because the master integral in ${\cal I}_{111}$ cannot be ignored, since it contributes in the limit in which {\it all} the frequencies go to zero ${\omega_i \to 0}$ yielding, as we shall see, a nonlocal-in-time {\it boundary} term. We discuss all the relevant memory corrections below. 

\subsubsection{Local-in-time}\label{sec:locM}

For the local-in-time part of the memory we follow the procedure in \cite{Blumlein:2021txe,Almeida:2022jrv} (but with the corrected Feynman rules), and ignore the contribution from ${\cal I}_{111}$. We find ($\omega_2 \equiv \omega_3 - \omega_1$),
\begin{align}
	S^{\textrm{(e)}}_{\rm loc\textrm{(M)}} & = \frac{G^2}{35}\int \frac{\dd \omega_3 \dd \omega_1}{(2\pi)^2} 
		I_{-}^{ij}(-\omega_3)I_{+}^{jk}(\omega_1)I_{+}^{ki}(\omega_2)
		\bigg[ 
		\omega_1^2\omega_2^2 \left(\omega_1^4+10 \omega_1^3\omega_2+25 \omega_1^2\omega_2^2+10 \omega_1\omega_2^3+\omega_2^4\right)
		 \notag \\
		& \qquad - 2 
		\omega_3^2 \omega_1^2 \left(\omega_3^4-10 \omega_3^3 \omega_1+25 \omega_3^2 \omega_1^2-10 \omega_3  \omega_1^3+\omega_1^4\right) \bigg] \,,\\
	S^{\textrm{(f)\&(g)}}_{\textrm{(M)}} & = -\frac{G^2}{5} \int \frac{\dd \omega_3 \dd \omega_1}{(2\pi)^2} 
		I_{-}^{ij}(-\omega_3)I_{+}^{jk}(\omega_1)I_{+}^{ki}(\omega_2)
		\omega_1^3\left[ 
		2 \omega_3^5 -5 \omega_3^4 \omega_1 +2 \omega_3^3 \omega_1^2  \right.\notag \\ &\qquad+ \left.\frac{\omega_2^3}{2}\big(2\omega_1^2+5 \omega_1\omega_2 +2\omega_2^2 \big)
		\right] \, .
\end{align}
Up to total time derivatives, these can be rewritten as\begin{align}
	S^{\textrm{(e)}}_{\rm loc \textrm{(M)}} & = \frac{G^2}{5} \int \dd t \left[
		 I_{-, ij} I\ord{4}_{+, jk} I\ord{4}_{+, ki} 
		-2 I\ord{4}_{-, ij} I\ord{4}_{+, jk} I_{+, ki} 
		+ \frac{8}{7} I\ord{2}_{-, ij} I\ord{3}_{+, jk} I\ord{3}_{+, ki}
		- \frac{12}{7} I\ord{3}_{-, ij} I\ord{3}_{+, jk} I\ord{2}_{+, ki} \right] \, , 
		\label{eq:Smema} \\
			S^{\textrm{(f)\&(g)}}_{\textrm{(M)}} & = \frac{G^2}{5} \int \dd t \left[
	I\ord{4}_{-, ij} I\ord{4}_{+, jk} I_{+, ki}  -
	\frac{1}{2} I_{-, ij} I\ord{4}_{+, jk} I\ord{4}_{+, ki}  -
	 I\ord{2}_{-, ij} I\ord{3}_{+, jk} I\ord{3}_{+, ki} +
	 2I\ord{3}_{-, ij} I\ord{3}_{+, jk} I\ord{2}_{+, ki}  \right] \, . 
	\label{eq:Smembc}
\end{align}
While the result in \eqref{eq:Smema} agrees with the previous computations in \cite{Blumlein:2021txe,Almeida:2022jrv}, due to a missing term from the tetrad field in the Feynman rules, the result in~\eqref{eq:Smembc} fails to match the double-bubble contribution(s). Putting both terms together, we finally arrive at
\begin{equation}
	S_{\rm loc\textrm{(M)}} = \frac{G^2}{5} \int \dd t \left[
	\frac{1}{2} I_{-, ij} I\ord{4}_{+, jk} I\ord{4}_{+, ki} 
	- I\ord{4}_{-, ij} I\ord{4}_{+, jk} I_{+, ki} 
	+ \frac{1}{7} I\ord{2}_{-, ij} I\ord{3}_{+, jk} I\ord{3}_{+, ki}
	+ \frac{2}{7} I\ord{3}_{-, ij} I\ord{3}_{+, jk} I\ord{2}_{+, ki}
	\right]\,.
	\label{eq:Mloc}
\end{equation}

\subsubsection{Soft-frequency limit} 
In order to capture the soft-frequency limit (SFL) contribution to the memory diagram in~\ref{Fig:Sinin}(e), it is instructive to rederive the integrand in terms of the stress-energy tensor, prior to integrating over the $\bar h_\pm$ field(s), which can be split as (see also \cite{Almeida:2023yia} for the case of tail terms)
\begin{align}
\label{cut111}
	  \textrm{Fig}.\,\ref{Fig:Sinin}{\rm (e)} &=  \int \frac{\dd \omega_1 \dd \omega_2 \dd \omega_3}{(2\pi)^3}\ddl(\omega_1 + \omega_2 +\omega_3) \notag \\
	& \int_{\bk} \bigg[ 
	\frac{-i \kappa}{2} \Tmempom 
	\bigg]^{\mu\nu}
	\frac{i P_{\mu\nu\rho\sigma}}{(\omega_1+\omega_2+i0^+)^2-\bk^2} 
	\bigg[ 
	\frac{-i \kappa}{2} \Tzeromeqom \bigg]^{\rho\sigma}  + {\rm perms.}\,,
\end{align}%
where the permutations account for the various possible {\it cuts} of the {\it would-be} ${\cal I}_{111}$ (modulo the proper symmetry factors).  Moreover, since the soft limit is dominated by three radiation modes, we can further consider the associated {\it cut} propagators to be on-shell, which greatly simplifies the computation.
Let us start with the first term displayed in \eqref{cut111} and consider the limit $\omega_1+\omega_2 \to 0$, for which we get
\begin{equation}
	\bigg[
	\lim_{\omega_1+\omega_2 \to 0}\Tmempom
	\bigg]^{\mu\nu} = \frac{1}{\omega_1+\omega_2+i0^+}T^{\mu\nu}_{\textrm{SFL(M)}}(\omega_1+\omega_2, \bk) \, .
\end{equation}
such that, 
\begin{align}
	 \textrm{Fig}.\,\ref{Fig:Sinin}{\rm (e)}\bigg|_{\omega_1+\omega_2 \to 0} & = -i\frac{\kappa^2}{4} \int \frac{\dd \omega_1 \dd \omega_2 \dd \omega_3}{(2\pi)^3}\ddl(\omega_1 + \omega_2 +\omega_3)\frac{1}{\omega_1+\omega_2+i0^+} \notag \\
	& \times \int_\bk T^{\mu\nu}_{+\textrm{SFL(M)}}(\omega_1+\omega_2, \bk)\frac{P_{\mu\nu\rho\sigma}}{(-\omega_3+i0^+)^2-\bk^2} T_{-,(LO)}^{\rho\sigma}(\omega_3, \bk) \, .
\end{align}
After tensorial reduction, and solving the integral over $\bk$, we find 
\begin{align}
	\textrm{Fig}.\,\ref{Fig:Sinin}{\rm (e)}\bigg|_{\omega_1+\omega_2 \to 0} \hspace{-0.3cm} = i\frac{2 G^2}{35} \int \frac{\dd \omega_1 \dd \omega_2 \dd \omega_3}{(2\pi)^3}
	I_+^{ij}(\omega_1)I_+^{jk}(\omega_2)I_-^{ki}(\omega_3)\frac{\omega_3^3\omega_1^6}{\omega_1+\omega_2+i0^+}
	\ddl(\omega_1 + \omega_2 +\omega_3) \,.
	\label{eq:fig4ZFL}
\end{align}
Hence, using the distributional identity in~\eqref{eq:PVplusdelta}, and keeping only the nonlocal-in-time term, in which all the frequencies go to zero, we arrive at 
\begin{align}
	&\textrm{Fig}.\,\ref{Fig:Sinin}{\rm (e)}\bigg|_{\omega_1+\omega_2 \to 0, \omega_3 \to 0} \hspace{-0.2cm}  = \frac{G^2}{35}\int \frac{\dd \omega_1 \dd \omega_2 \dd \omega_3}{(2\pi)^3}\omega_1^6\omega_3^3 \ddl(\omega_1+\omega_2)\ddl(\omega_1+\omega_2+\omega_3) I_+^{ij}(\omega_1)I_+^{jk}(\omega_2)I_-^{ki}(\omega_3) \, .
	\label{eq:deltafig4}
\end{align}
The contribution from the other cuts involves the coupling of $T^{\mu\nu}_{+\rm (LO)}$ with $T^{\mu\nu}_{-\rm SFL(M)}$ instead. However, since the diagram turns out to be proportional to two terms (with retarded and advanced propagators) of opposite signs, we find that the latter vanishes in the soft-frequency limit.  Hence, we arrive at the final result
\begin{align}
	S_{\rm nloc(M)} = -i\frac{G^2}{35}\bigg[ & \int \frac{\dd \omega_1 \dd \omega_3}{(2\pi)^2} \ddl(\omega_3)\omega_3^3 I_-^{ij}(\omega_3)  \omega_1^6 I_+^{jk}(-\omega_1)I_+^{ki}(\omega_1) \bigg] \label{eq:111} \, ,
\end{align}
proportional to $\omega^3 \delta(\omega) I^{ij}(\omega)$.\footnote{At the level of the  ${\cal I}_{111}$ integral, the term in \eqref{eq:111} would appear, for instance, from a contribution proportional to (schematically)
\beq
\begin{aligned}
&\,\,\int\frac{\dd\omega_{1,2,3}}{(2\pi)^3}\,\omega_1^6 \mathscr{M}_+(\omega_1)\mathscr{M}_+(\omega_2) \mathscr{M}_-(-\omega_3) (\omega_1+\omega_2-\omega_3)^2 \delta(\omega_1+\omega_2-\omega_3) \left( \frac{\omega_3}{\omega_1+\omega_2+i0^+}+\cdots\right)  \\  
&\, =-\frac{i}{2}\int\frac{\dd\omega_{1,3}}{(2\pi)^2}\,\omega_1^6 \omega_3^3 \ddl(\omega_3) \mathscr{M}(\omega_1)_+\mathscr{M}_+(-\omega_1) \mathscr{M}_-(\omega_3) +\cdots \,.\nn
\end{aligned}
\eeq
}
It is somewhat instructive to regroup some of the terms together, and rewrite the entire contribution from the soft-frequency limit as follows
\begin{equation}
	S_{\textrm{SFL(M)}}= -\frac{2 G^2}{35} \int \!\!\dd t \bigg [ 
	 I_+^{(6)ij}(t)I_+^{jk}(t)I_-^{(2)ki}(t) - \frac{\dd}{\dd t} \bigg\{ I_-^{(2)ij}(t) \int \dd \tau \vartheta(t - \tau)I_+^{(6)jk}(\tau)I_+^{ki}(\tau) \bigg\}\bigg ]\label{eq:nlocsflm}\,,	
	\end{equation}
after absorbing the associated local-in-time contribution. Notice, as anticipated, the nonlocal-in-time part becomes a total time derivative, hence a boundary term in the action that does not explicitly contribute to the equations of motion. 
\subsection{Radiation-reaction forces} \label{sec:Force}

The total in-in effective action, including radiative effects, takes the general form,
\begin{equation}
	S[\bx_{\pm}] = \int \dd t \Big( \mathcal{L}+ R_{\pm} \Big)  \, ,
\end{equation}
where $\mathcal{L}$ is a conservative-like Lagrangian, including  the kinetic and potential-only terms computed to 5PN order in \cite{Blumlein:2020pyo}, as well as other conservative-like tail terms, e.g.~\cite{Galley:2015kus,Almeida:2021xwn}. On~the other hand, the $R_{\pm}$ part accounts for various radiation-reaction effects, starting with the leading (Burke-Thorne) radiation-reaction action \cite{Galley:2015kus}
\beq
\label{eq:srr}
S_{\rm (RR)} = -\frac{G}{5} \int \dd t \, I^{ij}_{-}(t) I^{ij(5)}_+(t)\,.
\eeq
The equation of motion are then given by \cite{Galley:2009px}
\begin{equation}
\label{eq:eom}
	\frac{\dd}{\dd t}\frac{\partial \mathcal{L}}{\partial \bv^i_a} - \frac{\partial \mathcal{L}}{\partial \bx^i_a} = \bigg[\frac{\partial R_{\pm}}{\partial \bx^i_{a,-}} 
		-\frac{\dd}{\dd t}\frac{\partial R_{\pm}}{\partial \bv^i_{a,-}} \bigg]\bigg|_{\textrm{PL}} \,,
\end{equation}
where PL stands for ``Physical Limit", i.e., $\bx_{a, +} \to \bx_a$, $\bx_{a, -} \to 0$.\vskip 4pt At this stage we treat the full failed-tail and memory contributions as part of the radiative sector, and discuss their conservative-like counterparts shortly. Using \eqref{eq:eom} we obtain\footnote{Notice that, as expected, the ${\cal O}(G^3)$ terms are (Schott-type) total time derivatives.}
\begin{align}
	\ba_{(\textrm{FT})} & = \frac{G^3 M^3 \nu^2}{r^4}\bigg[
	\bigg(
	36 v^6 - 288 v^4 (\bv\cdot \bn)^2 + 112 v^2 (\bv\cdot \bn)^4 + 252 (\bv\cdot \bn)^6
	\bigg) \bn \notag \\
	& \qquad\qquad - \bigg(
	84 v^4 (\bv\cdot \bn) - 588 v^2 (\bv\cdot \bn)^3 + 616 (\bv\cdot \bn)^5 
	\bigg)
	\bv
	\bigg] \notag \\
	& -
	\frac{4G^4 M^4 \nu^2}{5r^5} \bigg[
	\bigg(
	47 v^4  + 163 v^2 (\bv\cdot \bn)^2 - 630 (\bv\cdot \bn)^4
	\bigg) \bn  - \bigg(
	350 v^2 (\bv\cdot \bn) - 770 (\bv\cdot \bn)^3 
	\bigg)
	\bv
	\bigg] \notag \\
	& +\frac{8G^5 M^5 \nu^2}{15r^6} \bigg[
	\bigg(
	11 v^2  + 199 (\bv\cdot \bn)^2 \bigg) \bn 
	- 210 (\bv\cdot \bn) \bv
	\bigg]  \, , \label{eq:FTrela} \\
		\ba_{(\textrm{M})} & = -\frac{2 G^3 M^3 \nu^2}{35 r^4}\bigg[
	\bigg(
	2484 v^6 - 38325 v^4 (\bv\cdot \bn)^2 + 91770 v^2 (\bv\cdot \bn)^4 - 56385 (\bv\cdot \bn)^6
	\bigg) \bn \notag \\
	& \qquad\qquad + \bigg(
	9633 v^4 (\bv\cdot \bn) - 33870 v^2 (\bv\cdot \bn)^3 + 24885 (\bv\cdot \bn)^5 
	\bigg)
	\bv
	\bigg] \notag \\
	& +
	\frac{4G^4 M^4 \nu^2}{315r^5} \bigg[
	\bigg(
	13900 v^4  - 95892 v^2 (\bv\cdot \bn)^2 + 76680 (\bv\cdot \bn)^4
	\bigg) \bn  \notag \\
	& \hspace{2cm}+ \bigg(
	32163 v^2 (\bv\cdot \bn) - 35955 (\bv\cdot \bn)^3 
	\bigg)
	\bv
	\bigg] \notag \\
	& -\frac{16G^5 M^5 \nu^2}{1260 r^6} \bigg[
	\bigg(
	2028 v^2  + 9421 (\bv\cdot \bn)^2 \bigg) \bn 
	- 393 (\bv\cdot \bn) \bv
	\bigg]  
	-\frac{824G^6 M^6 \nu^2}{63 r^7} \bn\, , \label{eq:Mrela} 
\end{align}
for the relative acceleration(s), $\ba \equiv \ba_1-\ba_2$, with $M=m_1+m_2$, $\nu = \frac{m_1m_2}{M^2}$, $\br \equiv \bx_1-\bx_2$, $\bn \equiv \br/r$, and $\bv \equiv \bv_1-\bv_2$. Since these will be useful later on, we also quote the Burke-Thorne radiation-reaction force, both after inputing the Newtonian acceleration, \beq \label{eq:N} \ba_{\rm (N)} = -GM\frac{\br}{r^3}\,,\eeq
at linear order,
\begin{align}
	\ba_{\rm (RR)} & =  \frac{8 G^2 M^2 \nu}{5 r^3}\bigg[
	\bigg(18 v^2-25 (\bv\cdot \bn)^2 \bigg) (\bv\cdot \bn) \bn
	-\bigg(6 v^2 - 15(\bv\cdot \bn)^2
	\bigg) \bv
	\bigg]  + \notag \\
	&  \hspace{2cm} + \frac{16 G^3 M^3 \nu}{5 r^4} \bigg[
	\bv +\frac{\bv\cdot \bn}{3} \bn
	\bigg] \, ,
	\label{eq:BTaLO} 
\end{align}
and at second order, after plugging it back onto itself,
\begin{align}
\label{eq:BTaNLO}
\ba_{\rm (RR^2)}& =  \frac{16 G^3 M^3 \nu^2}{5 r^4}\bigg[
	\bigg(
	168 v^6-3465 (\bv \cdot \bn)^6 -2496 v^4 (\bv \cdot \bn)^2+5789 v^2 (\bv \cdot \bn)^4
	\bigg)  \bn \\
	& \hspace{2cm} +\bigg(
	738 v^4+1715 (\bv \cdot \bn)^4-2449 v^2 (\bv \cdot \bn)^2
	\bigg) (\bv \cdot \bn)\bv
	\bigg]  \notag \\
	& - \frac{32 G^4 M^4\nu^2}{225 r^5}\bigg[
	\bigg(
	3898 v^4-28131 v^2 (\bv \cdot \bn)^2+22755 (\bv \cdot \bn)^4
	\bigg)\bn \notag \\
	& \hspace{2cm} +6 \bigg(
	1921 v^2-1830 (\bv \cdot \bn)^2
	\bigg)(\bv \cdot \bn) \bv
	\bigg] \notag \\
	& + \frac{64 G^5 M^5 \nu^2}{225 r^6}\bigg[
	\bigg(
	311 v^2 + 689(\bv \cdot \bn)^2
	\bigg)\bn
	-234(\bv \cdot \bn) \bv
	\bigg]
	+\frac{1792 G^6 M^6 \nu^2}{75 r^7} \bn\, . \notag
\end{align}
 
\subsection{Near-zone dissipation}\label{sec:Flux}

We are now in position to check that the near-zone dynamics is consistent with the radiated energy and angular momentum computed from the metric perturbation at infinity. Because of the subtleties with the extra boundary term in the action, we split the discussion into local- and nonlocal-in-time terms.  
\subsubsection{Local-in-time}

We start by constructing the energy and angular momentum of the two-body system through the (local-in-time) conservative part of the action, 
\begin{align}
	M \equiv \sum_{a=1,2} \bv_a \cdot \frac{\partial \mathcal{L}}{\partial \bv_a} - \mathcal{L} \,  , \qquad \frac{1}{2}L^{ij} \equiv \sum_{a=1,2} \bx_a^{[i} \frac{\partial \mathcal{L}}{\partial \bv_a^{j]}}\,,
\end{align}
such that the Euler-Lagrangian equations imply 
\begin{align}
	\dot{M} & = \sum_a \bv_a^\ell \bigg[\frac{\partial R_{\pm}}{\partial \bx^\ell_{a,-}} 
		-\frac{\dd}{\dd t}\frac{\partial R_{\pm}}{\partial \bv^\ell_{a,-}} \bigg]\bigg|_{\textrm{PL}} \, , 
		\label{eq:MdotR} \\
	\dot{\bL}_{i} & = \sum_a \varepsilon_{i j k} \bx_a^{j} \bigg[\frac{\partial R_{\pm}}{\partial \bx^{k}_{a,-}} 
		-\frac{\dd}{\dd t}\frac{\partial R_{\pm}}{\partial \bv^{k}_{a,-}} \bigg]\bigg|_{\textrm{PL}} \, ,
		\label{eq:LdotR}
\end{align}
where we keep the full failed-tail and memory radiation-reaction forces on the right-hand-side of these equations.\footnote{In principle, some of these terms may belong to the conservative sector, turning into total derivatives that can then be moved to the left-hand-side.} Performing various manipulations (see App.~\ref{app:MdotLdot}) we find from the failed-tail forces the (averaged) mass/energy loss %
\begin{equation}
	\langle \dot{M} \rangle_\textrm{(FT)} = - \frac{G^2}{15} \bigg \langle L_{kl} I\ord{1}_{i k}(t)I\ord{7}_{l i}(t)\bigg\rangle + \frac{G^2}{30} \bigg\langle \dot{L}_{kl} I_{i k}(t)I\ord{7}_{l i}(t)\bigg\rangle \, .
\end{equation}
Since, at this order in the perturbative expansion we have $\dot{L}_{ij} = 0 + {\cal O}(G^2)$, the second term can be ignored so that \begin{equation}
	\langle \dot{M} \rangle_\textrm{(FT)} = \frac{G^2}{15}\bigg\langle L_{kl} I\ord{4}_{i k}(t)I\ord{4}_{l i}(t) \bigg\rangle+ \left\langle \frac{\dd}{\dd t}( \cdots )\right\rangle = 0 + {\cal O}(\dot{L}_{ij})\, ,
\end{equation}
which is then consistent with~\eqref{eq:MdotflFT}. Following similar steps for the angular momentum loss, 
\begin{equation}
	\langle\dot{\bL}_{n}\rangle_\textrm{(FT)} =-\frac{G^2}{15}\left\langle\varepsilon_{nij} I_{ik}\Big(
	L_{kl}I\ord{7}_{lj} + L_{jl}I\ord{7}_{kl}
	\Big)\right\rangle +
	\frac{G^2}{30}\left\langle\varepsilon_{nij} L_{ik} \Big(
	I_{jl}I\ord{7}_{kl} -  I_{kl}I\ord{7}_{jl} 
	\Big)\right\rangle \, ,
\end{equation}
and using that 
\begin{equation}
	\varepsilon_{nij}L_{kl}I_{ik}I\ord{7}_{lj} = \frac{1}{2}\varepsilon_{nij}L_{kl} \Big(
	I_{ik}I\ord{7}_{lj} + I_{lj}I\ord{7}_{ki}
	 \Big) = 0 +\frac{\dd}{\dd t}(\cdots) + {\cal O}(\dot{L}_{ij}) \,,
\end{equation}
we are left with the second term, such that 
\begin{equation}
	\langle\dot{\bL}_{n}\rangle_\textrm{(FT)} = \frac{G^2}{30}\varepsilon_{nij}\left\langle\Big( 
	-3 I\ord{7}_{kl} I_{ik}  + I_{kl} I\ord{7}_{ik} 
	\Big)L_{jl}\right\rangle = \frac{2 G^2}{15} \varepsilon_{nij} \left\langle I\ord{4}_{kl}I\ord{3}_{k i}L_{j l}\right\rangle + {\cal O}(\dot{L}_{ij}) \, ,
\end{equation}
which is also consistent with~\eqref{eq:LdotFT}. Let us stress, as we mentioned before, that the consistency between results requires, crucially, the inclusion of the term proportional to $L_{-}^{ij}$ in \eqref{eq:Sfail}, and its associated radiation-reaction force.\vskip 4pt

The computation of the radiated energy and angular momentum for the local-in-time part of the memory follows the same steps. The derivation of the mass/energy loss is straightforward, and we find 
\begin{align}
	 \langle \dot{M}\rangle_\textrm{(M)} &  = \frac{G^2}{5}\left \langle I\ord{1}_{ij} I\ord{4}_{jk} I\ord{4}_{ki} \right\rangle +\left\langle \frac{\dd}{\dd t}(\cdots)\right\rangle  \, ,
\end{align}
consistent with \eqref{eq:Mdotmemory}, whereas for the angular momentum loss, we arrive at
\begin{align}
	\langle\dot{\bL}_{k}\rangle_{\textrm{loc(M)}} & = -\frac{2G^2}{5} \varepsilon_{kij} \left\langle I\ord{4}_{ai}(t) I_{j b}(t)I\ord{4}_{ab}(t)\right\rangle \,,
\end{align}
that is also consistent with~\eqref{eq:dotLML}.

\subsubsection{Boundary term}

The contribution from the nonlocal-in-time part is somewhat subtle. First of all, since it is a total derivative, it does not partake in the near-zone equations of motion directly. However, it does affect the definition of the {\it conserved} charges. Following~\cite{Galley:2009px,Galley:2014wla}, we can find the value of the linear and angular momentum, as well as the energy, by varying the (on-shell) nonlocal-in-time effective action (after integrating by parts)\beq
S_{\rm nloc (M)}(T,\bx_{a,\pm}(T)) = - \frac{4 G^2}{35}  \sum_a m_a \left[\left( \bv^i_{a,-} (T)\bv^j_{a,+}(T)\right)_{\rm STF} \int_{-\infty}^T I_+^{(3)jk}(\tau)I^{(3)ik}_+(\tau) \dd \tau\right]\,,
\eeq 
with respect to the end points at $T \to + \infty$. Notice we have only kept the (nonvanishing) term proportional to $ \ddot I^{ij} \sim  \big(\bv^i \bv^j\big)_{\rm STF}$. The nonlocal-in-time contribution to the (canonical) angular momentum is then obtained from the variation
\beq
\bL^\ell_{\rm nloc (M)}(T) = \varepsilon^{\ell mn}\sum_a \bx^m_{a,+}(T) \frac{\partial S_{\rm nloc (M)}}{\partial \bx_{a,-}^n(T)}\bigg|_{\textrm{PL}}\,.
\eeq
Hence, using that as $T \to +\infty$ the (local-in-time) evolution equations become $\bx(T) \to \bv(T) T$, such that $\bv(T) = \bx(T)/T$, we find
\beq
\bL^\ell_{\rm nloc (M)}(T \to +\infty)  =  -\frac{4 G^2}{35} \lim_{T \to +\infty} \varepsilon^{\ell mn} \left[ I^{(2)mj}(T) \int_{-\infty}^T I^{(3)n\ell}(\tau)I^{(3)k\ell}(\tau) \dd \tau\right]\,,
\eeq  
such that the total change of angular momentum will coincide with the integral over the flux in \eqref{eq:LdotFluxmem}. We can also perform similar steps for the other charges.  For instance, for the energy, we can easily show that
\beq
\begin{aligned}
&\qquad \quad E_{\rm nloc(M)} (T \to +\infty) =   -  \lim_{T \to +\infty} \frac{\partial S_{\rm nloc (M)}}{\partial T}  = \\ &-8 \lim_{T \to +\infty} \sum_a m_a \left\langle \bv^i_{a} (T)\ba^j_{a}(T)\right \rangle_{\rm STF} \int_{-\infty}^T I_+^{(3)jk}(\tau)I^{(3)ik}_+(\tau) \dd \tau \\ 
&-4 \lim_{T \to +\infty} \sum_a m_a \left\langle \bv^i_{a} (T)\bv^j_{a}(T)\right \rangle_{\rm STF}I_+^{(3)jk}(T)I^{(3)ik}_+(T) \dd \tau  \to 0 \,, 
\end{aligned}
\eeq
Similarly, the nonlocal-in-time contribution to the linear momentum vanishes at infinity. Notably, the angular momentum involves the product $|\br||\bp| \simeq T/T$, which remains finite as we take $T \to + \infty$. Let us emphasize that the loss of (canonical) angular momentum is directly associated with Noether's charge, i.e. $\br \times \bp$, while the particle's angular momentum, i.e.~$\br\times \bv$, is not affected by the nonlocal-in-time boundary term.\footnote{This is not at all surprising and happens also, for instance, in electromagnetism with a constant (fixed) vector-potential, $\bA$, which induces a coupling $\tfrac{\dd}{\dd t}(\br \cdot \bA)$ in the worldline's action. Although the particle's Lagrangian is not invariant under rotations, which implies $\tfrac{\dd}{\dd t}(\br\times \bp) = \bv \times \bA \neq 0$, the breaking is due to a total derivative. Hence, the action is invariant and $\br \times \bv$ is conserved instead. Let us  retain the same coupling and assign now different initial/final values to the vector-potential at infinity. Similarly to the gravitational memory, but for the linear (canonical) momentum, we find $\Delta \bp \propto \Delta \bA$. Yet, since the extra term is still a total derivative, the particle's momentum, $m\bv$, remains constant.}  We will explore the implications of the nonlocal-in-time memory effect in more detail elsewhere. 

\section{Conservative action}\label{sec:cons}
A conservative-like contribution (in the sense introduced in \cite{Galley:2009px,Galley:2014wla}), can be obtained by using the in-out boundary conditions with Feynman propagators and retaining the real part of the answer~\cite{Kalin:2022hph}. This procedure was utilized in \cite{Bern:2021yeh,Dlapa:2021vgp,Dlapa:2022lmu,Dlapa:2023hsl,Dlapa:2024cje,Driesse:2024xad} and is also consistent with the TF approach at ${\cal O}(G^4)$ \cite{Bini:2021gat}. Hereditary terms, however, are not the end of the road and radiation-reaction-square effects can also introduce conservative-like corrections. This is, in fact, also the case of electromagnetic radiation, for which we provide in App.~\ref{RRQED} an explicit derivation, in agreement in the overlap with the recent relativistic results obtained in~\cite{Bern:2023ccb}. The analogous contributions in gravity are due to nonlinear effects in the Burke-Thorne force \cite{Burke:1970dnm,Galley:2009px}, which we discuss below. In the next section we will incorporate all of the nonlinear gravitational effects to derive of the conservative part of the scattering angle at ${\cal O}(G^4)$. 

\subsection{Feynman's prescription}

For the derivation of the in-out effective action we must introduce Feynman's boundary conditions, using 
\beq
\begin{aligned}
i\Delta_F(\tau,\bx=0) &= -\int \frac{\dd\omega}{2\pi} \int_{\bk} \frac{ e^{-i \omega \tau }
}{\omega^2 -\bk^2+ i 0} = \frac{i}{4\pi} \int \frac{\dd\omega}{2\pi}|\omega|e^{-i \omega \tau} = \frac{i}{4\pi^2}\partial_\tau \frac{\PV}{\tau}\label{absw}\,,
\end{aligned}
\eeq 
instead of retarded propagators, where $\PV$ stands for the Principal-Value distribution, also known as `Hilbert transform,'
\begin{equation}
	 \int \frac{\dd \omega}{2\pi} (i \textrm{Sign}(\omega) ) e^{-i\omega\tau} = \frac{1}{\pi} \frac{\PV}{\tau} \, .
\end{equation}
For hereditary effects, as well as at second-order in the radiation reaction force, after IBP relations are implemented we will encounter products of (at most) two Feynman propagators. For the case of tail and failed-tail terms, the frequency of the radiation field is not modified by the interaction with the background geometry and we may simply replace the resulting factor of $|\omega|^2$ by $-(i\omega)^2$, and trade it for time derivatives. However, for memory contributions, new subtleties arise, since we will often find products of the sort, $|\omega_1| |\omega_2|$, involving two different frequencies. For instance, even ignoring the subtle soft-frequency limits we discussed before,  the conservative effective action will then include terms of the form (schematically)
\begin{align}
 S^{\rm cons}_{\rm (M)} \supset \int \dd t_1 \,    \int \dd t_2 \frac{\PV}{t_1-t_2} \int \dd t_3 \frac{\PV}{t_2-t_3}  \mathscr{M}_1(t_1) \mathscr{M}_2(t_2)  \mathscr{M}_3 (t_3)\,,\label{eq:Smem} 
\end{align}
where the $\mathscr{M}_{1,2,3}$ depend on derivatives of the quadrupole moment, producing an effective action with a novel nonlocal-in-time dependence.\vskip 4pt

In order to deal with the above expression, we will evaluate the effective action perturbatively in Newton's constant. First of all, we notice that whenever we have a contribution from $\mathscr{M}_{1(0)}(t_1)$ or $\mathscr{M}_{3(0)}(t_3)$, evaluated on an {\it unperturbed} solution at ${\cal O}(G^0)$, the associated $\PV$-integral vanishes. The situation is more subtle for the $\mathscr{M}_{2(0)}(t_2)$ term, which shares the common $dt_2$ integral over the principal values. In that case, it is useful to invoke the Poincar\'e-Bertrand theorem, as a distributional identity \cite{Davies:1990fe,Davies:1996gee},
\begin{equation}
	\int \dd x \frac{\PV}{t - x} \int \dd y \frac{\PV}{x - y} = \int \dd y \int \dd x \bigg[
	\frac{\PV}{(t-x)(x - y)} - \pi^2 \delta(t-x)\delta(x-y)
	\bigg]
	\, ,\label{pbthm}
\end{equation}
to exchange the order of integration, yielding
\beq
\begin{aligned}
 S^{\rm cons}_{\rm (M)} \supset 
 & - \pi^2 \int \dd t \, \mathscr{M}_1(t) \mathscr{M}_{2(0)}(t)\mathscr{M}_3(t) \,,
\label{pbthm2} 
 \end{aligned}
 \eeq 
 after the integral over the double principal value now vanishes, reducing to a similar expression as tail-type terms.\vskip 4pt  Following the above procedure at all orders in $G$, we can always separate conservative memory effects into local- and nonlocal-in-time terms.\footnote{Notice that, unlike the claims in \cite{Blumlein:2021txe} (where the $\PV$-integrals are de facto ignored), the analytic properties of the quadrupole moment in situations of interest will in general produce a nontrivial Hilbert transform.} It is then straightforward to show that the latter only start to contribute at 5PM order, and therefore it will not affect the comparison with the conservative 4PM results in \cite{Dlapa:2021vgp,Bern:2021yeh} (see below). We will return to this issue in \S\ref{sec:disc}, and in more detail elsewhere.
   
\subsection{Failed tail \& Memories} 

The Feynman diagrams for the in-out computation are identical to the in-in version, except for symmetry factors and the replacement of retarded by Feynman propagators. Combining all the pieces, we have for the failed tail, 
\begin{equation}
	S^{\rm cons}_{\textrm{(FT)}} = +\frac{1}{15} \frac{i\kappa^4 }{128}\int \frac{\dd\omega\dd q_0}{(2\pi)^2}\int_{\bk, \bq} \frac{L^{kl}(q_0) \omega^5 I^{ik}(-\omega)I^{il}(\omega -q_0) }{[\omega^2 - \bk^2 +i0^+] [(\omega-q_0)^2 - \bq^2 +i0^+]} \, .
\end{equation}
Notice that the static nature of the total angular momentum sets $q_0 \to 0$, such that the relevant Feynman integral becomes
\begin{equation}
	\int_{\bk, \bq} \frac{1}{[\omega^2 - \bk^2 +i0^+] [\omega^2 - \bq^2 +i0^+]} = -\frac{\omega^2}{16\pi^2} \, ,
\end{equation}
with the same frequency on both propagators. Hence,  
\begin{equation}
\label{SconsFT}
	S^{\rm cons}_{\textrm{(FT)}}  = -\frac{G^2}{30}\int \dd t \,L^{kl} I^{(4) i k} I^{(3)il} \, ,
\end{equation}
which is in agreement with the previous derivations in \cite{Almeida:2023yia,Henry:2023sdy}.\vskip 4pt 

Following similar steps, and ignoring the soft-frequency contribution, the derivation of the in-out memory integrand is straightforward, and we arrive at
 \begin{align}
 	S^{\rm cons}_{\rm (M)} & = \frac{8 G^2 \pi^2}{35} \int \frac{\dd \omega \dd \omega_1}{(2\pi)^2}\int_{\bq,\bk}\frac{f(\omega,\omega_1)}{[\omega^2 - \bk^2 +i0^+] [\omega_1^2 - \bq^2 +i0^+]}  I^{ij}(\omega_1) I^{jk}(\omega-\omega_1) I^{ki}(-\omega)\,\nn\\
	&=  -\frac{G^2 }{70} \int \frac{\dd \omega \dd \omega_1}{(2\pi)^2}|\omega| |\omega_1| f(\omega,\omega_1) I^{ij}(\omega_1) I^{jk}(\omega-\omega_1) I^{ki}(-\omega)\,, \label{SconsM}
	\end{align}
where
\beq
f(\omega,\omega_1) \equiv \omega  \omega_1 \left(2 \omega ^4 -6 \omega ^3 \omega_1+15 \omega ^2 \omega_1^2 - 6 \omega  \omega_1^3+2 \omega_1^4\right)\,,
\eeq
featuring, as we mentioned, the product of absolute values of two independent frequencies. At 4PM order, however, it is straightforward to show that one of the multipole moments must always be {\it static}. We then notice that whenever we take the unperturbed solution for $I_{(0)}^{ij}(\omega_1)$, or $I_{(0)}^{ki}(\omega)$, the integral over the frequency vanishes. That is the case either due to high powers of the frequency or an integral that is odd under parity. (Notice the latter holds only because of the absolute value.) The surviving term is thus proportional to $I_{(0)}^{jk}(\omega-\omega_1) \propto \delta^{(n)}(\omega-\omega_1)$, with $n=0,1,2$, in which case the action becomes (as in \eqref{pbthm2}), 
\beq
 \begin{aligned}
 \label{SconsM2}
	S^{{\rm cons}}_{G^4 \rm (M)} & = -\frac{ G^2}{70} \int \frac{\dd \omega \dd \omega_1}{(2\pi)^2} \omega^2  \omega_1^2 \left(2 \omega ^4 -6 \omega ^3 \omega_1+15 \omega ^2 \omega_1^2 \right.\\ &\left.\qquad\qquad - 6 \omega  \omega_1^3+2 \omega_1^4\right) I_{\rm (1)}^{ij}(\omega_1) I_{(0)}^{jk}(\omega-\omega_1) I_{(1)}^{ki}(-\omega)\, \\
	&= \frac{G^2}{5} \int \dd t \bigg(
	\frac{1}{7}I^{(2)ij} I^{(3)jk} I^{(3)ik} 
	-\frac{1}{2}I^{ij} I^{(4)jk} I^{(4)ik}\bigg)_{G^2}\,,
	\end{aligned}
\eeq
where $ I_{\rm (1)}^{ij} \propto G^1$ is evaluated on the Newtonian solution, and we have rearranged the terms in the second line such that this is manifest. (Notice that the result in \eqref{SconsM2} differs from the one obtained by ignoring the absolute values in \eqref{SconsM}.)  At the end of the day, conservative memory effects at ${\cal O}(G^4)$ then resemble a (local-in-time) tail-type contribution with a (static) quadrupolar interaction, which is also consistent with the type of integrals appearing in the PMEFT derivation in~\cite{Dlapa:2021vgp}. 

\subsection{Radiation-reaction square}

In order to incorporate the remaining radiation-reaction-square effects, we start by inserting the (imaginary) linear force onto itself, thus producing a conservative (real) part of the total radiative result.\footnote{Technically speaking, this is done by starting from the equations of motion with retarded Green's functions, and splitting the latter into Feynman plus a {\it reactive} term. The (real part of the) product of (two) Feynman propagators yields a conservative contribution that coincides with the product of (two) time-symmetric Green's functions.} 
Similarly to what we find in the case of electromagnetism (see App.~\ref{RRQED}), the conservative part arises from the following (real) effective action, 
\beq
S^{\rm cons}_{\rm (RR^2)} = \frac{i}{10\pi} \int \dd t_1 \dd t_2 I^{ij}(t_1) \frac{\PV}{t_1-t_2} I^{(5)ij}_{\rm (RR)}(t_2)\,, \label{eq:rr21}
\eeq
where
\begin{align}
	I\ord{5}_{ij\rm (RR)}(t_2) & = i \frac{4G}{5\pi} L_{k \langle i}(t_2)\int \dd t_3 \frac{\PV}{t_2 - t_3}I\ord{7}_{j\rangle k}(t_3) + i\frac{16 G}{5\pi} Q\ord{1}_{k \langle i}(t_2)\int \dd t_3 \frac{\PV}{t_2 - t_3}I\ord{7}_{j\rangle k}(t_3) \notag \\
	& \quad + i \frac{20  G}{5\pi} Q\ord{2}_{k \langle i}(t_2)\int \dd t_3 \frac{\PV}{t_2 - t_3}I\ord{6}_{j\rangle k}(t_3) 
	 + i\frac{8 G}{5\pi} Q\ord{3}_{k \langle i}(t_2)\int \dd t_3 \frac{\PV}{t_2 - t_3}I\ord{5}_{j\rangle k}(t_3) \notag \\
	& 
	\quad + i \frac{4 G}{5\pi} Q_{k \langle i}(t_2)\int \dd t_3 \frac{\PV}{t_2 - t_3}I\ord{8}_{j\rangle k}(t_3) \,,
\end{align}
after replacing $L^{ij} = 2\sum_a m_a \bx_a^{[i} \bv_a^{j]}$ and we have introduced $Q^{ij} \equiv \sum_a m_a \bx_a^i\bx_a^j$ (which includes also the traces that are absent in $I^{ij}$). Hence, plugging it back into \eqref{eq:rr21} and using the fact that $\dot{L}_{ij} = 0$  at this order, we find%
\beq
\begin{aligned}
	S^{\rm cons}_{(\textrm{RR}^2)} = -\frac{2G^2}{25 \pi^2}& \int \dd t_1 \dd t_2 \frac{\PV}{t_1 - t_2} I_{ij}(t_1)\bigg\{
	L_{k \langle i}(t_2)\int \dd t_3 \frac{\PV}{t_2 - t_3}I\ord{7}_{j\rangle k}(t_3)  \\
	& + 4Q\ord{1}_{k \langle i}(t_2)\int \dd t_3 \frac{\PV}{t_2 - t_3}I\ord{7}_{j\rangle k}(t_3) 
	+ 5Q\ord{2}_{k \langle i}(t_2)\int \dd t_3 \frac{\PV}{t_2 - t_3}I\ord{6}_{j\rangle k}(t_3) \\
	& + 2Q\ord{3}_{k \langle i}(t_2)\int \dd t_3 \frac{\PV}{t_2 - t_3}I\ord{5}_{j\rangle k}(t_3) 
	+ Q_{k \langle i}(t_2)\int \dd t_3 \frac{\PV}{t_2 - t_3}I\ord{8}_{j\rangle k}(t_3)
	\bigg\} \,.
\end{aligned}
\eeq
In order to extract the ${\cal O}(G^4)$ contribution, we rewrite the above expression in Fourier space and notice, as before, that only the term proportional to $Q_{(0)}^{ij}(\omega-\omega_1)$ survives, yielding 
\beq
\begin{aligned}
	S^{\rm cons}_{G^4 (\textrm{RR}^2)} & = \frac{2 G^2}{25}  \int \frac{\dd \omega \dd q_0}{(2\pi)^2} (-i\omega^7)L^{ki}(q_0)I^{kj}(-\omega)I^{ij}(\omega -q_0)
	 \\
	& +\frac{2 G^2}{25} \int \frac{\dd \omega \dd \omega_1}{(2\pi)^2}\bigg\{
	\Big( 2\omega_1^3 \omega^5 -\omega_1^2 \omega^6 \Big)I^{ij}(\omega_1) Q_{(0)}^{ki}(\omega-\omega_1)I^{jk}(-\omega) 
	\bigg\} \, \\
	& = \frac{2 G^2}{25}\int \dd t \left(-  L_{ki} I\ord{4}_{kj}I\ord{3}_{ji} + Q_{ki}
	 I\ord{4}_{kj}I\ord{4}_{ij} + Q\ord{2}_{ki} I\ord{3}_{kj}I\ord{3}_{ij}\right)_{G^2}\label{SconsRR2}\,. 
\end{aligned}
\eeq 
This action can now be used alongside the failed-tail and memory results to describe the conservative dynamics at 5PN/4PM order, and in particular to derive the conservative scattering angle as we show momentarily.

\section{Scattering data at 5PN/4PM order}\label{sec:angle}

We are now in position to combine our results together with the other, potential-only plus tail-type, contributions to the (in-in) effective theory and compute the total (relative) scattering angle at 5PN order. We will focus in the overlap between the (even-in-velocity) hereditary plus radiation-reaction-square corrections and the total 4PM results in \cite{Dlapa:2022lmu}. as well as the value in \cite{Dlapa:2021vgp,Bern:2021yeh} for the conservative part, respectively. We derive the corrections due to the failed-tail, memory, as well as radiation-reaction-square terms, by integrating their contribution to the relative impulse on the trajectories,
\begin{equation}
	\Delta \bp = M \nu \int \dd t \bigg( \ba_{(\textrm{N})}(t)+ \ba_{\textrm{(RR)}}(t) + \ba_{\rm (RR^2)}(t) + \ba_{\textrm{(FT)}}(t)+ \ba_{\textrm{(M)}}(t) \bigg) \,,
		\label{eq:relpgen}
\end{equation}
to the desired 5PN/4PM order. We write the resulting relative angle in terms of a PM expansion, 
\beq
\frac{\chi_{\rm rel}}{2} = \sum_n \chi_{b,\rm rel}^{(n)}(v_{\infty})  \left(\frac{GM}{b}\right)^n = \sum_n \chi_{j,\rm rel}^{(n)} (v_{\infty}) \frac{1}{j^n}\,,
\eeq
where $b \equiv |\bb|$, with $\bb$ the impact parameter,  $j = \frac{J}{GM\mu}$ the (reduced) angular momentum, and $v_{\infty}$ the (relative) velocity at infinity, such that $\chi_{j,\rm rel}^{(n)} = \left(\frac{v_\infty}{\Gamma}\right)^n \chi_{b,\rm rel}^{(n)}$, with $\Gamma \equiv \sqrt{1+2\nu(\gamma-1)}$ and $\gamma \equiv \sqrt{1+v_\infty^2}$.  For reasons that will become clear momentarily, we will also introduce the variable \cite{Bini:2021gat} \beq \tilde \chi_{j,\rm rel}^{(n,\nu^2)} \equiv \Gamma^{n-1}\chi_{j,\rm rel}^{(n,\nu^2)}\,,\eeq 
retaining only the ${\cal O}(\nu^2)$ part.\footnote{The contributions at ${\cal O}(\nu)$ are already in perfect agreement between all the existent literature.} To alleviate notation, we will also drop the `rel' tag on the angle from now on.

\subsection{Potential-only \& Tails}

The contribution from generic tail effects to the in-in effective action has been computed within the EFT approach in \cite{Almeida:2021xwn}, yielding 
\beq
\begin{aligned}\label{tails}
S_{I_2\rm (T)} &= \frac{2G^2 M_+}{5}\int \frac{\dd\omega}{2\pi} \omega^6 I_-^{ij}(-\omega)I_+^{ij}(\omega)  \left[
	-\frac{1}{d-3} +\frac{41}{30} + i \pi\, \textrm{sign}(\omega) - \log \left(\frac{\omega^2e^{\gamma_{\textrm{E}}}}{\pi\mu^2}\right)
	\right] \, , \\
S_{I_3\rm (T)} &= \frac{2G^2 M_+}{189}\int \frac{\dd\omega}{2\pi} \omega^8  I_-^{ijk}(-\omega)I_+^{ijk}(\omega)  \left[
	-\frac{1}{d-3} + \frac{82}{35} + i \pi \,\textrm{sign}(\omega) - \log \left(\frac{\omega^2e^{\gamma_{\textrm{E}}}}{\pi\mu^2}\right)
	\right] \, , \\
S_{J_2\rm(T)} &= \frac{32G^2 M_+}{90}\int \frac{\dd\omega}{2\pi} \omega^6  J_-^{a|ij}(-\omega)J_+^{a|ij}(\omega) \left[
	-\frac{1}{d-3} + \frac{49}{20} + i \pi \,\textrm{sign}(\omega) - \log \left(\frac{\omega^2 e^{\gamma_{\textrm{E}}} }{\pi\mu^2}\right)
	\right] \, ,
\end{aligned}
\eeq
for the terms that are relevant for our purposes here (see \cite{Galley:2015kus,Almeida:2021xwn,Amalberti:2023ohj} for more details). As it is well known,  the $1/(d-3)$ pole and factor of $\log\mu$ cancels out against the contribution from potential modes \cite{Galley:2015kus,Porto:2017dgs,Foffa:2019yfl,Blumlein:2020pyo} to 5PN order, yielding finite and ambiguity-free results.\vskip 4pt From the effective action we then obtain the equations and motion. For the conservative part, one can instead simply evaluate the conservative action (which follows ignoring the $i\pi\,\textrm{sign}(\omega)$ and dividing by a factor of two \cite{Galley:2015kus}) and taking a derivative with respect to the angular momentum. This was performed in \cite{Bini:2021gat} using previous EFT results, obtaining 
\begin{align}
\tilde \chi_{j\rm (pot+T)}^{(4,\nu^2) \rm cons} &= -\frac{8919}{1400} \pi \nu^2 v_\infty^6\,.\label{chiconspott}
\end{align}
For the dissipative part, on the other hand, we must compute the impulse from the radiative force, involving the factors of $i\pi\,\textrm{sign}(\omega)$ in \eqref{tails}. As it is well known, these term reproduce the total radiated energy due to tail effects \cite{Galley:2015kus,Bini:2021gat}. Furthermore, their contribution(s) to the (relative) scattering angle may be also obtained by using linear-response theory. Following the analysis in \cite{Bini:2021gat}, we find 
\begin{align}
\tilde \chi_{j\rm (T)}^{(4) \rm diss} &= 0 \,.\label{chitdis}
\end{align}
The vanishing of the dissipative part of the tail term at ${\cal O}(G^4)$ turns out to have important consequences regarding the origin of various different contributions to the near-zone dynamics. 
 
\subsection{Failed tail, Memories \& RR$^2$}

We start with the failed-tail and memory parts, which follows via \eqref{eq:relpgen}
evaluated on the deflected trajectories,
\begin{equation}
	\br(t) = \br_0(t)+ \delta_{\textrm{(FT)}}\br(t) + \delta_{\textrm{(M)}}\br(t) \,,
\end{equation}
where $\br_0 = \bb + \bv_\infty t $, and the perturbed trajectories, $\delta_{\textrm{(FT)}}\br(t)$ and $\delta_{\textrm{(M)}}\br(t)$, satisfy %
\begin{equation}
\label{eq:deltax}
	\delta_{(\textrm{X})}\ddot{\br}(t) = \ba_{\textrm{(X)}}(t) \, , 
	\qquad \textrm{X} \in \{\textrm{FT, M}\} \, .
\end{equation}
Since the additional nonlocal-in-time memory term does not contribute to the total impulse, we may ignore it in what follows.\vskip 4pt  
We find, for each, two contributions to the total impulse. Firstly, we have the hereditary radiation-reaction force evaluated up to the Newtonian deflection, 
\begin{align}
	\Delta^{(\textrm{N})} \bp_{(\textrm{FT})} =  M \nu \int \dd t\, \ba_{\textrm{(FT)}} \Big|_{\br_0(t) + \delta_{(\textrm{N})}\br(t)} = -\frac{11}{16}\pi \frac{G^4 M^5 \nu^3}{b^4}v_\infty^3\frac{\bb}{b} +{\cal O}(G^5) \, ,\nn \\
	\Delta^{(\textrm{N})} \bp_{(\textrm{M})} =  M \nu \int \dd t \, \ba_{\textrm{(M)}} \Big|_{\br_0(t) + \delta_{(\textrm{N})}\br(t)} = \frac{871}{2240}\pi \frac{G^4 M^5 \nu^3}{b^4}v_\infty^3\frac{\bb}{b} +{\cal O}(G^5) \, .
\end{align}
The second part is obtained by including $\delta_{\textrm{(X)}}\br(t)$ in the Newtonian acceleration, 
\begin{align}
	\Delta^{(\textrm{FT})} \bp_{(\textrm{N})} =  M \nu \int \dd t\, \ba_{\textrm{(N)}} \Big|_{\br_0(t) + \delta_{(\textrm{N})}\br(t)+\delta_{(\textrm{FT})}\br(t)} = -\frac{35}{16}\pi \frac{G^4 M^5 \nu^3}{b^4}v_\infty^3\frac{\bb}{b}  + {\cal O}(G^5)\, ,\nn \\
	\Delta^{(\textrm{M})} \bp_{(\textrm{N})} =  M \nu \int \dd t \,\ba_{\textrm{(N)}} \Big|_{\br_0(t) + \delta_{(\textrm{N})}\br(t)+\delta_{(\textrm{M})}\br(t)} = \frac{7593}{2240}\pi \frac{G^4 M^5 \nu^3}{b^4}v_\infty^3\frac{\bb}{b}  + {\cal O}(G^5)\,.
\end{align}
Combing all contributions we arrive at%
\begin{align}
	 \Delta^{\textrm{(N)}} \bp_{(\textrm{FT})} + \Delta^{(\textrm{FT})} \bp_{(\textrm{N})} = -\frac{23}{8}\pi \frac{G^4 M^5 \nu^3}{b^4}v_\infty^3\frac{\bb}{b}  +{\cal O}(G^5) \, ,\\
 \Delta^{(\textrm{N})} \bp_{(\textrm{M})} + \Delta^{(\textrm{M})} \bp_{(\textrm{N})} = \frac{529}{140}\pi \frac{G^4 M^5 \nu^3}{b^4}v_\infty^3\frac{\bb}{b}  + {\cal O}(G^5)\,.
\end{align}

For the remaining contribution at second order in the radiation-reaction force, we perform the same steps, but in this case we must evaluate \eqref{eq:relpgen} on trajectories obeying an equation as in \eqref{eq:deltax} with $X \in \rm \{N,RR,RR^2\}$ terms given by the accelerations in \eqref{eq:BTaLO} and \eqref{eq:BTaNLO}.\footnote{Let us remind the reader that, even though  $\ba_{\rm (RR^2)}$ is a Schott-type term at ${\cal O}(G^3)$, it enters in the impulse at ${\cal O}(G^4)$ through the deflected trajectory on the Newtonian force.} 
Using the same notation as before we find the following intermediate contributions,
\begin{equation}
	\Delta^{\rm (N)} \bp_{\rm (RR^2)} =  M \nu \int \dd t\, \ba_{\rm (RR^2)} \Big|_{\br_0(t) + \delta_{(\textrm{N})}\br(t)} = \frac{3567}{200}\pi \frac{G^4 M^5 \nu^3}{b^4}v_\infty^3\frac{\bb}{b}  + {\cal O}(G^5)\, ,
\end{equation}
\begin{align}
	\Delta^{\rm (RR)} \bp_{\rm (RR)} &=  M \nu \int \dd t \,\ba_{\rm (RR)} \Big|_{\br_0(t) + \delta_{(\textrm{N})}\br(t)+\delta_{(\textrm{RR})}\br(t)} = -\frac{479}{25}\pi \frac{G^4 M^5 \nu^3}{b^4}v_\infty^3\frac{\bb}{b}  +{\cal O}(G^5)\,,\\

	\Delta^{(\textrm{RR}^2)} \bp_{(\textrm{N})} &=  M \nu \int \dd t \, \ba_{\textrm{(N)}} \Big|_{\br_0(t) + \delta_{(\textrm{N})}\br(t)+\delta_{(\textrm{RR})}\br(t)+\delta_{(\textrm{RR}^2)}\br(t)} = -\frac{791}{40}\pi \frac{G^4 M^5 \nu^3}{b^4}v_\infty^3\frac{\bb}{b}  +{\cal O}(G^5) \, , 
\end{align}
which combines into
\begin{equation}
	\Delta^{\rm (N)} \bp_{\rm (RR^2)}+\Delta^{\rm (RR)} \bp_{\rm (RR)} + \Delta^{(\textrm{RR}^2)} \bp_{(\textrm{N})} = -\frac{211}{10}\pi \frac{G^4 M^5 \nu^3}{b^4}v_\infty^3\frac{\bb}{b} +{\cal O}(G^5)  \, . \label{dprr2}
\end{equation}
Although we disagree with the previous hereditary (failed-tail and memory) values computed in \cite{Almeida:2022jrv}, the expression in \eqref{dprr2} is consistent with their combined result for radiation-reaction-square terms. 

\subsection{Total deflection angle}

Adding up the results from nonlinear radiation-reaction effects we obtain the total impulse, 
%
from which we can derive the scattering angle, via
\begin{equation}
	\chi = \arccos\left( \frac{\bp_- \cdot \bp_+}{|\bp_-||\bp_+|} \right) \, ,
\end{equation}
with the $\bp_\pm$ the incoming and outgoing relative momenta, respectively. Using the standard convention that the deflection is positive along the $-\hat\bb$ direction (as the leading order), the individual contributions coming from failed-tail, memory and (combined) radiation-reaction-square effects, are given by
\beq
\begin{aligned}
	\tilde \chi^{(4,\nu^2)}_{j(\rm FT)} & = \frac{23}{16}\pi \nu^2 v_\infty^6 \,,\\
	\tilde \chi^{(4,\nu^2)}_{j(\rm M)} & = -\frac{529}{280}\pi \nu^2 v_\infty^6\,, \\
	\tilde \chi^{(4,\nu^2)}_{j({\rm RR}^2)} & = \frac{211}{20} \pi \nu^2 v_\infty^6\,,
\end{aligned}
\eeq
respectively. Summing these results together with the total potential and tail terms in~\eqref{chiconspott} (and 
\eqref{chitdis}), we finally arrive at
\begin{align}
	\tilde \chi^{(4,\nu^2)\rm tot}_{j(\rm even)} & = \frac{1491}{400} \pi \nu^2 v_\infty^6 \,,\label{eq:tot4pmangle}
\end{align}
for the even-in-velocity ${\cal O}(G^4)$ coefficient of the total (relative) scattering angle, entering at second order in the mass ratio. This result is in perfect agreement in the overlap with the value computed in~\cite{Dlapa:2022lmu} at 4PM order. 
\subsection{Conservative part}\label{sec:locons}

Using the expressions in \eqref{SconsFT}, \eqref{SconsM2} and \eqref{SconsRR2} for the hereditary and radiation-reaction-square contributions to the effective action, respectively, we can readily derive the associated correction to the conservative deflection angle at ${\cal O}(G^4)$, by evaluating the (radial) action on the Newtonian trajectory to the given order in $G$, and taking a derivative with respect to the angular momentum (see App.~\ref{RRQED}). We find 
\beq
\begin{aligned}
	\tilde \chi^{(4,\nu^2)\rm cons}_{j(\rm FT)} & = \frac{69}{80}\pi \nu^2 v_\infty^6 \,,\\
	\tilde \chi^{(4,\nu^2)\rm cons}_{j(\rm M)} & = -\frac{477}{560}\pi \nu^2 v_\infty^6\,, \\
	\tilde \chi^{(4,\nu^2)\rm cons}_{j({\rm RR}^2)} & = \frac{159}{25} \pi \nu^2 v_\infty^6\,,
\end{aligned}
\eeq
which, together with~\eqref{chiconspott}, yields
\beq
\tilde \chi^{(4,\nu^2)\rm cons}_{j \rm{(pot+T+FT+M+RR^2)}} = 
\bigg(-\frac{8919}{1400} + \frac{69}{80} -\frac{477}{560} + \frac{159}{25}
\bigg)\pi \nu^2 v_\infty^6 = 0\,,\label{chimass0}
\eeq
in perfect agreement in the overlap with the conservative 4PM result in \cite{Dlapa:2021vgp,Bern:2021yeh}, as well as the value inferred from the TF formalism in \cite{Bini:2021gat}, consistently with the expected (polynomial) mass-scaling of the scattering computation \cite{Damour:2019lcq} (see also \cite{Vines:2018gqi,Kalin:2019rwq}).

\section{Conclusions \& Outlook}\label{sec:disc}

In this paper we have computed the missing hereditary effects in the near-zone two-body (relative) dynamics, yielding the following result for the complete (in-in) effective action to 5PN order,
\beq
\begin{aligned}	
\label{eq:fin} 
	&S^{\rm tot}_{\rm 5PN} =  S_{\rm (pot)}+S_{
	\rm (RR)} + S_{\rm (T)}  -\frac{G^2}{15}\int \! \dd t \, L_{+}^{kl}I\ord{4}_{-,kj}I\ord{3}_{+, jl}
		 + \frac{G^2}{30}\int \! \dd t \, L_{-}^{kl} I\ord{4}_{+,kj}I\ord{3}_{+, jl}  \\
		& + \frac{G^2}{5} \int \dd t \left(
	\frac{1}{2} I_{-, ij} I\ord{4}_{+, jk} I\ord{4}_{+, ki} 
	- I\ord{4}_{-, ij} I\ord{4}_{+, jk} I_{+, ki} 
	+ \frac{1}{7} I\ord{2}_{-, ij} I\ord{3}_{+, jk} I\ord{3}_{+, ki}
	+ \frac{2}{7} I\ord{3}_{-, ij} I\ord{3}_{+, jk} I\ord{2}_{+, ki} 
	\right)  \\
	&- \frac{2G^2}{35} \int \!\!\dd t 
 \frac{\dd}{\dd t} \left\{ I_-^{(2)ij}(t) \int \dd \tau \vartheta(t - \tau)I_+^{(3)jk}(\tau)I_+^{(3)ki}(\tau)\right\}\,,
\end{aligned}
\eeq
where $S_{\rm (pot)}$ are the potential-only contributions obtained in \cite{Blumlein:2020pyo}, and the linear radiation-reaction, $S_{\rm (RR)}$, and tail terms, $S_{\rm (T)}$, are given in \eqref{eq:srr} and \eqref{tails}, respectively. \vskip 4pt 

Our results differ from previous derivations in several notable ways. First of all, there is a crucial term (depending on $L^{ij}_-$) in the failed-tail contribution which was not included before. Secondly, due to the enforcing of diffemorphism invariance in the Feynman rules, the coefficients of the (local-in-time) memory contributions also differ with the values in \cite{Blumlein:2021txe,Almeida:2022jrv}. Finally, because of the nontrivial soft-frequency limit of the relevant Feynman integrals, we have uncovered a novel nonlocal-in-time (boundary) term. From the expression in \eqref{eq:fin} we derived the equations of motions, and demonstrated the consistency between the near- and far-zone GW fluxes, paying particular attention to the connection between boundary terms and the flux of (canonical) angular momentum. From the equations of motion we computed the contribution to the total (even-in-velocity) relative scattering angle at ${\cal O}(G^4)$, including the potential-only, tail, and radiation-reaction-square terms, finding complete agreement in the overlap with the state of the art in the PM expansion~\cite{Dlapa:2022lmu}.\vskip 4pt

We have also discussed the split into conservative and dissipative parts, following Feynman's prescription. We have found several subtleties due to the introduction of $\PV$-integrals. Moreover, we demonstrated the presence of radiation-reaction-square terms in the conservative sector. After adding all the relevant pieces, the conservative dynamics at 5PN/4PM may be obtained from the following effective action, 
\beq
\begin{aligned}
\label{Sconstot4}
 S^{\rm cons}_{\rm 5PN/4PM}&=  S_{\rm (pot)} + S^{\rm cons}_{\rm (T)} + \frac{G^2}{5}\int \dd t \bigg\{-\frac{1}{6} \,L^{kl} I^{(4)ki}I^{(3)li}+
 \\ &
	\frac{1}{7}I^{(2)ij} I^{(3)jk} I^{(3)ik} -\frac{1}{2}I^{ij} I^{(4)jk} I^{(4)ik}+\\
	& \frac{2}{5} \left(-  L^{ki} I^{(4)kj}I^{(3)ij} + Q^{ki}
	 I^{kj(4)}I^{(4)ij} + Q^{(2)ki} I^{(3)kj}I^{(3)ij}\right)	\bigg\}\,,
	 \end{aligned}
	 \eeq
expanded to ${\cal O}(G^4)$, where $S^{\rm cons}_{\rm(T)}$ are the conservative parts of tail terms \cite{Galley:2015kus,Almeida:2021xwn}, and the above expression incorporates the remaining failed-tail, memory, and radiation-reaction-square effects, respectively. Notice that the last line introduces terms that resemble the others, as well as dependence on traces that are absent from hereditary effects. From here we derived the conservative scattering angle, finding also perfect agreement with the PN-exact ${\cal O}(G^4)$ value obtained in \cite{Dlapa:2021vgp,Bern:2021yeh}, as well as the 5PN/4PM result inferred from the TF approach~\cite{Bini:2021gat}.\vskip 4pt  As it was discussed in \cite{Dlapa:2024cje}, upon subtracting nonlocal-in-time~tail effects (from the $\log\omega^2$ in \eqref{tails}), the resulting scattering angle may be analytically continued through the boundary-to-bound dictionary \cite{Kalin:2019rwq,Kalin:2019inp,Cho:2021arx} to compute observables for generic bound orbits, or incorporated into a local-in-time Hamiltonian (see also~\cite{Bini:2024tft}). Likewise, at higher orders in $G$, the expression in \eqref{Sconstot4} will incorporate all of the ``tail-like" 5PN contributions to the conservative sector, provided the multipole moments with $n \leq 2$ derivatives are kept unperturbed (setting the acceleration to zero). However, starting at 5PM, other conservative (Feynman) memory effects---depending on an integral over the principal value---will introduce nonlocal-in-time terms that are not captured by~\eqref{Sconstot4}. We will return to a more in-depth discussion of the conservative sector and the connection to scattering elsewhere. \vskip 4pt
 
There are various other aspects of our calculation that have uncovered somewhat unexpected issues that deserve further study:
\begin{itemize}
\item  \textbf{\textit{Tails vs Memories}}

As~we discovered, the 4PM dissipative contribution, denoted as $\chi_{b,\rm rel}^{(4)\rm 2rad}$ in \cite{Dlapa:2022lmu}, is entirely captured by failed-tail, memory, and radiation-reaction-square terms in the PN EFT approach. Indeed, the overall scaling with the mass already implied that this term had to originate from the product of three quadrupole moments (albeit one of them is {\it static}, as in tail-like interactions).\footnote{This is consistent with the fact that $\chi_{b,\rm rel}^{(4)\rm 2rad}$ can be obtained by combining the tail-induced radiated linear momentum (see e.g. Eq. (H1) in \cite{Bini:2021gat}) together with the relations uncovered in Eqs. (12.34-12.36) of \cite{Bini:2022enm}.} The situation gets more interesting at higher orders in $G$. Starting at 5PM, regions with {\it three} radiation modes will contribute in relativistic scattering computations (such as tail-of-tail effects). Yet, after IBP reduction we find that (modulo soft-frequency limits) all of the leading memory effects reduce to master integrals with {\it two} radiation modes (not necessarily with the same frequency). This apparent dichotomy is remediated by noticing that the time derivatives in hereditary terms can themselves be evaluated on the (linear) radiation-reaction force. Moreover, the three-bubble diagram (as well as the tail of the memory) will also enter at ${\cal O}(G^5)$, evaluated on the Newtonian solution. These considerations illustrate the intricate connections between the multipole expansion, the method of regions, and the various contributions from nonlinear interactions in both PM and PN effective theories. 

\item  \textbf{\textit{ Conservative-like RR$^2$}} 

Conservative effects entailed also terms at second order in the linear radiation-reaction force. At first sight, the Burke-Thorne force is purely dissipative, which is manifest in the fact that Feynman's computation is purely imaginary, and so is the associated radiation-reacted trajectory. The conservative part of the force arises after incorporating the radiation-reaction acceleration onto itself. As in the case of electromagnetism, this was sufficient at ${\cal O}(G^4)$. However, at higher orders,  conservative-like contributions to the impulse may arise from iterations over the radiation-reacted trajectory itself. Yet, this cannot be inferred from the iteration over a real conservative force, which implies that one has to be careful when splitting conservative and dissipative radiation-reaction-square terms. In particular, to avoid double counting, not only we ought to isolate the conservative part of $\ba_{\rm (RR^2)}$, but in principle also the iterations over $\ba_{\rm (RR)}$.

\item  \textbf{\textit{Conservative (time) nonlocality}}

 As we have shown, despite Feynman's prescription introducing new non-analyticities in the frequencies, the latter do not play a role in the derivation of conservative effects at ${\cal O}(G^4)$, which may be obtained directly from the effective action in~\eqref{Sconstot4}, having only the already known time nonlocality due to tail effects \cite{Galley:2015kus}. However, starting at ${\cal O}(G^5)$ Feynman's prescription may incur in additional nonlocal-in-time (memory and radiation-reaction-square) effects, proportional to a $\PV$-integral.~Yet, since the total result derived from \eqref{eq:fin} is local, any extra nonlocality will cancel out against a counterpart in the dissipative sector. This suggests that we can perform a splitting of the (relative) dynamics into local-in-time conservative and dissipative terms (for instance by constructing time-symmetric expressions in terms of retarded and advanced propagators). However, we cannot assume the former will always obey the same (polynomial) mass scaling inherited from Feynman's computation, as in \eqref{chimass0}. This implies that we may not be able to simply compare PN and PM results beyond 4PM order, and only total values may be immune to different choices. 
 
 \item  \textbf{\textit{Soft-frequency limit}}

Another subtle issue is the soft-frequency limit of the Feynman integral involving the cubic coupling. Even though it does not affect the impulse, it does contribute to the flux of (canonical) angular momentum. This is not entirely surprising, since it is well known that the angular-momentum flux is sensitive to the waveform in the $\omega \to 0$ limit, see e.g.~\cite{Damour:2020tta,Manohar:2022dea,DiVecchia:2022owy,Riva:2023xxm}. Moreover, we found that the soft-frequency contribution due to memory effects in~\eqref{eq:dotLMNL} starts at ${\cal O}(G^5)$, which is also consistent with the results in \cite{Bini:2021qvf,Heissenberg:2024umh}. However, there are two key aspects regarding the derivation of the near-zone action. Firstly, the need to avoid enforcing the energy/frequency conservation prior to performing the IBP decomposition; and secondly, the appearance of factors of $\tfrac{\omega^3}{\omega+i0}$ which cannot be naively simplified in the presence of a nontrivial analytic structure of the multipole moments. The soft-frequency-limit contribution enters as a boundary term, thus affecting the canonical momentum while the evolution equations for the position and velocities remain unaltered. This also suggests that the impact of the nonlocal-in-time angular-momentum flux on the evolution of the GW phase is more subtle than what we would have naively expected from the far-zone computation.

\end{itemize}

\vskip 4pt 

In addition to the above there are other directions worth of further exploration. As we discussed, we have concentrated on the relative near-zone dynamics.\footnote{As we mentioned, the center-of-mass recoil can be included through terms depending on the center-of-mass position and velocity in the effective action, yielding a correction to the radiation-reaction force, or directly via the flux of momentum using the Ward identity.} The issue with the relative part, for instance for the impulse, is that it does not capture (directly) the total radiated momentum \cite{Bini:2021gat}. The latter, however, is the only type of GW flux from failed-tail, memory, and radiation-reaction-square effects that does not turn into a total time derivative after writing the result in terms of positions and velocities (modulo the nonlocal-in-time flux of canonical angular momentum). Hence, these effects are {\it conservative} from the point of view of the relative dynamics, since they may be reabsorbed into the left-hand-side of a balance-type equation.~This implies, for instance, that the total (even-in-velocity) relative scattering angle due to nonlinear gravitational forces at 5PN order may be described in terms of a Hamiltonian, albeit without the mass-polynomiality of the Feynman result, as we see already in \eqref{eq:tot4pmangle}. We will return to this issue in more detail in forthcoming work.\vskip 4pt

Finally, there is the connection with the more traditional  Multipolar-Post-Minkowskian (MPM) formalism \cite{Blanchet:1985sp}. Although we have checked that the total radiated energy and angular-momentum agree with the MPM results in \cite{Bini:2022enm},\footnote{This follows form the fact that the (in-in) EFT and MPM fluxes agree up to Schott terms.} the explicit functional form of the memory part of the $h_{ij}^{\rm TT}$ one-point function---written in terms of products of derivatives of the quadrupole moment---does not formally agree with the MPM form~\cite{Arun:2009mc}, except for the nonlocal-in-time contribution. Given the notorious differences between both frameworks, this is somewhat expected. For starters, the couplings in the effective action do not include traces, while these are kept throughout the MPM approach. Furthermore, the EFT multipoles are defined with respect to a locally-flat frame and matched with the (pseudo-)stress-energy tensor; whereas in the MPM formalism other coordinates and matching conditions are used.
Yet, agreement has been found so far in all observable quantities, notably the recent rederivation in \cite{Amalberti:2024jaa} of the energy flux and radiated power at 3PN order, first obtained in the MPM approach \cite{Blanchet:2013haa}; as well as the rederivation in \cite{Marchand:2017pir} of 4PN conservative tail effects, first obtained within the EFT approach~\cite{Galley:2015kus}. Moreover, in the PM regime, the results in \cite{Bini:2022enm} are consistent with the total values in~\cite{Dlapa:2022lmu}, and recent MPM and amplitude-based approaches have produced matching results for the waveform in the overlapping realm of validity \cite{Bini:2024rsy,Georgoudis:2024pdz,Bini:2024ijq}. Hence, we expect that once all the pieces are collected (including terms at second order in the radiation-reaction, which are also formally different), the value of the full waveform in the EFT and MPM approaches, once written in terms of gauge-invariant quantities, will ultimately agree. We will return to the explicit comparison elsewhere. At the same time, we expect the MPM derivation of the 5PN near-zone dynamics will agree as well with the results reported here, which may help us elucidate the connection between formalisms.\footnote{Another possible route is to recast the EFT derivations in a way that resembles the MPM approach \cite{Riccardo}.}

\section*{Acknowledgments}  We thank Christoph Dlapa, Gregor K\"alin and Zhengwen Liu for discussions on the integration problem, and Donato Bini and Francois Larrouturou for discussions on the MPM formalism. The work presented here was supported by the ERC-CoG ``Precision Gravity: From the LHC to LISA" provided by the European Research Council under the European Union's H2020 research and innovation program (grant agreement No. 817791). MMR is also partially financed by the Deutsche Forschungsgemeinschaft (DFG) under Germany's Excellence Strategy (EXC 2121)
``Quantum Universe" (390833306). RAP would like to thank the International Center for Theoretical Physics - South American Institute for Fundamental Research (ICTP-SAIFR) and Instituto Principia  for hospitality while this paper was prepared for submission, as well as Alan M\"uller and Riccardo Sturani for discussions on related work.

\appendix

\section{Background gauge \& Ward identities}\label{app:WI}


Throughout this paper we have emphasized the importance of the background-gauge fixing in~\eqref{eq:BGGF} for diffeomorphism invariance and the validity of the Ward identities. However, for the computations we performed, the additional part of the Feynman rule due to the background-field gauge did not play a role. Needless to say, this will not be true in general. We show here an example where the background-field method is essential to guarantee that the associated $T^{\mu\nu}(x)$ satisfies the Ward identity.\vskip 4pt

\begin{figure}[t]
\begin{center}
	\includegraphics{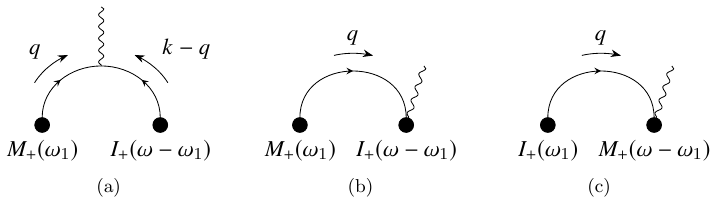}	
\end{center}
\caption{The  Feynman diagrams needed for the computation of the tail contribution to $T_+^{\mu\nu}$. For simplicity we denote $ k^0= \omega$ and  $q^0=\omega_1$. }
\label{Fig:Wtail}
\end{figure}

Let us consider the diagrams in Fig.~\ref{Fig:Wtail} responsible for the tail effects, where we included also the contributions which are needed for a nonstatic monopole term. Hence, we have a (Bondi) mass/energy coupling where the $M(\omega_1)$ is {\it not} proportional to $\delta(\omega_1)$ and include, for instance, corrections due to GW emission. In background gauge we have the gauge-fixing harmonic term\begin{equation}
	S_{\textrm{GF}} = \int \dd^4 x \sqrt{-\bar{g}} \bar{g}^{\mu\nu}
		\left[\bar{g}^{\alpha\beta}\bar{\nabla}_{\alpha} h_{\beta\mu} - 
		\frac{\bar{g}^{\alpha\beta}}{2}\bar{\nabla}_\mu h_{\alpha\beta}\right]
		\left[\bar{g}^{\rho\sigma}\bar{\nabla}_{\rho} h_{\sigma\nu} - 
		\frac{\bar{g}^{\rho\sigma}}{2}\bar{\nabla}_\nu h_{\rho\sigma}\right] \, ,
		\label{eq:GFBB2}
\end{equation} 
which, after expanding the background metric (with $\xi$ just a placeholder)%
\begin{equation}
	\bar{g}_{\mu\nu} = \eta_{\mu\nu} + \xi \kappa \bar{h}_{\mu\nu} \, ,
\end{equation}%
implies (schematically)
\begin{align}
	S_{\textrm{GF}} = \int \dd^4 x \bigg\{
	(\partial h)^2 + \kappa \xi \bigg[
	\bar{h} (\partial h)^2 + (\partial \bar{h})^2 h + (\partial \bar{h}) (\partial h) h 
	\bigg] + \cdots 
	\bigg\} \, ,
\end{align}
modifying the cubic interaction for the (``quantum") $h_{\mu\nu}$ field and the background $\bar{h}_{\mu\nu}$. Returning to the diagrams in~Fig.~\ref{Fig:Wtail}, we find that  Fig.~\ref{Fig:Wtail}(c) is manifestly zero, since it is always proportional to a scaleless integral which in dim. reg. vanishes. On the other hand, both  Fig.~\ref{Fig:Wtail}(a) and Fig.~\ref{Fig:Wtail}(b) contribute and, after an IBP reduction, the result can be written as 
\beq
\begin{aligned}
 T^{\mu\nu}_{+, \ref{Fig:Wtail}\textrm{(a)+(b)}}(k) & = \kappa^2  \int \frac{\dd \omega_1}{2\pi}  
 M_+(\omega) I_+^{\alpha\beta}(\omega - \omega_1)  \Big[ 
 \mathcal{N}\ord{1}_{\alpha\beta}{}^{\mu\nu}(\omega, \omega_1, \bk, \xi) \mathcal{I}_{1, 0}  \\
 & \hspace{1.2cm}
	+ \mathcal{N}\ord{2}_{\alpha\beta}{}^{\mu\nu}(\omega, \omega_1, \bk, \xi) \mathcal{I}_{0, 1} 
	+ \mathcal{N}\ord{3}_{\alpha\beta}{}^{\mu\nu}(\omega, \omega_1, \bk, \xi) \mathcal{I}_{1, 1} 
	\Big]
 \, ,	
\end{aligned}
\eeq
where we encounter the same master integrals we had before, i.e.
\begin{equation}
	\mathcal{I}_{ab} \equiv \int_{\bq} \frac{1}{\big[ (\omega_1 +i0^+)^2 -\bq^2 \big]^a \big[ (\omega - \omega_1 +i0^+)^2 -(\bk - \bq)^2 \big]^b} \, .
\end{equation}
Notice that all three tensorial structures $\mathcal{N}\ord{n}_{\alpha\beta}{}^{\mu\nu}(\omega, \omega_1, \bk, \xi)$ depend on the gauge-fixing choice in~\eqref{eq:GFBB2} through the $\xi$ parameter. In order to verify the Ward identity we then compute
\begin{align}
	 k_\nu T^{\mu\nu}_{+, {\rm Fig}.\ref{Fig:Wtail}\textrm{(a)(b)}}(k) & = 
	 \kappa^2  \int \frac{\dd \omega_1}{2\pi}  
 M_+(\omega) I_+^{\alpha\beta}(\omega - \omega_1)  \Big[ 
  k_\nu\mathcal{N}\ord{1}_{\alpha\beta}{}^{\mu\nu}(\omega, \omega_1, \bk, \xi) \mathcal{I}_{1, 0} \notag \\
 & \hspace{1.2cm}
	+  k_\nu\mathcal{N}\ord{2}_{\alpha\beta}{}^{\mu\nu}(\omega, \omega_1, \bk, \xi) \mathcal{I}_{0, 1} 
	+  k_\nu\mathcal{N}\ord{3}_{\alpha\beta}{}^{\mu\nu}(\omega, \omega_1, \bk, \xi) \mathcal{I}_{1, 1} 
	\Big] \notag \\
	& = \kappa^2 (1-\xi) \int \frac{\dd \omega_1}{2\pi}  
 \omega_1 M_+(\omega_1) I_+^{\alpha\beta}(\omega - \omega_1)  \Big[ 
  \tilde{\mathcal{N}}\ord{1}_{\alpha\beta}{}^{\mu}(\omega, \omega_1, \bk) \mathcal{I}_{1, 0} \notag \\
 & \hspace{1.2cm}
	+  \tilde{\mathcal{N}}\ord{2}_{\alpha\beta}{}^{\mu}(\omega, \omega_1, \bk) \mathcal{I}_{0, 1} 
	+  \tilde{\mathcal{N}}\ord{3}_{\alpha\beta}{}^{\mu}(\omega, \omega_1, \bk) \mathcal{I}_{1, 1} 
	\Big]
 \, ,	
 \label{eq:WIBG}
\end{align}
where $\tilde{\mathcal{N}}\ord{1}_{\alpha\beta}{}^{\mu}(\omega, \omega_1, \bk)$ is the resulting tensorial structure after contracting with $k^\nu$, and we have factored out the $\xi$ dependence already. First of all, we immediately notice that the Ward identity is automatically obeyed when $\dot M=0$, for which $\omega M(\omega) \propto \omega \delta(\omega) \to 0$. Secondly, for a nonstatic mass/energy coupling, the expression in \eqref{eq:WIBG} does not vanish unless we choose $\xi = 1$. Namely, the background gauge-fixing action in \eqref{eq:GFBB2}. Finally, let us stress that in this scenario the diagram in Fig.~\ref{Fig:Wtail}(b) is crucial to guarantee the validity of the Ward identity.

\section{Near-zone (local-in-time) GW fluxes}\label{app:MdotLdot}

From the in-in computation, we will generally find the following form of the `dissipative' part of the in-in effective action,
\begin{equation}
	R_\pm =  \sum_{\mathscr{M}}\mathscr{M}_-^L \mathscr{F}_+^L(\mathscr{M_+}) \, ,
\end{equation}
where $\mathscr{M}_-^L$ is a generic multipole, $\mathscr{F}_+^L(\mathscr{M_+})$ is a function of the $\mathscr{M_+}$ multipoles, both contracted over $L$ indices. From here we obtain the energy and angular momentum losses from the near-zone dynamics (omitting the sum over particles for simplicity) \begin{align}
	\dot{M} & = \bv\cdot\bigg[ \frac{\partial \mathscr{M}_-^L}{ \partial \bx_-}\mathscr{F}_+^L(\mathscr{M_+}) 
	- \frac{\dd}{\dd t} \Bigg(
	\frac{\partial \mathscr{M}_-^L}{ \partial \bv_-}\mathscr{F}_+^L(\mathscr{M_+})
	\Bigg) \bigg]\bigg|_{\textrm{PL}} \, , 
	\label{eq:Mdotgeneral}\\
	\dot{L}_{k} & = \varepsilon_{k i j} \bx_{i}\bigg[ \frac{\partial \mathscr{M}_-^L}{ \partial \bx^{j}_-}\mathscr{F}_+^L(\mathscr{M_+}) 
	- \frac{\dd}{\dd t} \Bigg(
	\frac{\partial \mathscr{M}_-^L}{ \partial \bv^{j}_-}\mathscr{F}_+^L(\mathscr{M_+})
	\Bigg) \bigg]\bigg|_{\textrm{PL}} \, . 
	\label{eq:Ldotgeneral}
\end{align}
Starting with~\eqref{eq:Mdotgeneral}, and upon time averaging over an orbit, we have \cite{Maia:2017gxn} 
\begin{equation}
	\langle \dot{M} \rangle =  \left\langle \bigg[
	\bv\cdot \frac{\partial \mathscr{M}_-^L}{ \partial \bx_-}
	+ \frac{\dd \bv}{\dd t} \cdot  
	\frac{\partial \mathscr{M}_-^L}{ \partial \bv_-} \bigg]
	\mathscr{F}_+^L(\mathscr{M_+})
	 \right\rangle\bigg|_{\textrm{PL}} \, .
\end{equation}
In order to evaluate the time derivatives, we use that
\begin{align}
	\frac{\partial \mathscr{M}_-^L}{ \partial \bx^\ell_-} \bigg|_{\textrm{PL}} & = 
	\bigg(\frac{\partial \bx_1^k}{ \partial x^\ell_-}\frac{\partial}{\partial \bx_1^k} + \frac{\partial \bx_2^k}{ \partial \bx^\ell_-}\frac{\partial}{\partial \bx_2^k} \bigg) 
	\Big( \mathscr{M}_1^L  - \mathscr{M}_2^L\Big)\bigg|_{\textrm{PL}} \notag \\
	& = \bigg(
	\frac{\delta^k{}_\ell}{2}\frac{\partial \mathscr{M}_1^L}{\partial \bx_1^k}
	-\frac{(-\delta^k{}_\ell)}{2}\frac{\partial \mathscr{M}_2^L}{\partial \bx_2^k}
	 \bigg)\bigg|_{\textrm{PL}} = \frac{\partial \mathscr{M}^L}{\partial \bx^\ell} \, ,
\end{align}
and similarly for the derivative with respect to the velocity. Hence, 
since the multipole moments do not depend explicitly on time (for a binary in isolation), we have %
\begin{equation}
	\frac{\dd}{\dd t} \mathscr{M}^L (\bx(t), \bv(t)) = \bv\cdot \frac{\partial \mathscr{M}^L}{ \partial \bx}
	+ \frac{\dd \bv}{\dd t} \cdot  
	\frac{\partial \mathscr{M}^L}{ \partial \bv} \, ,
\end{equation}
such that
\begin{equation}
	\langle \dot{M} \rangle = \left\langle\bigg[
	\bv\cdot \frac{\partial \mathscr{M}^L}{ \partial \bx}
	+ \frac{\dd \bv}{\dd t} \cdot  
	\frac{\partial \mathscr{M}^L}{ \partial \bv} \bigg]
	\mathscr{F}^L(\mathscr{M})
	 \right\rangle = \left \langle \frac{\dd \mathscr{M}^L}{\dd t} \mathscr{F}^L(\mathscr{M})\right\rangle \, .
\end{equation}

The computation of the angular-momentum flux is a bit trickier. Let us restrict the manipulations here local-in-time effects. We use the expression in~\eqref{eq:Ldotgeneral} and, upon time-averaging, we get
\begin{equation}
	 \langle\dot{\bL}_{k} \rangle = \varepsilon_{kij}\bigg\langle\bigg[
	\bx_{i} \frac{\partial \mathscr{M}_-^L}{ \partial \bx^{j}_-} 
	+ \bv_{i} 
	\frac{\partial \mathscr{M}_-^L}{ \partial \bv^{j}_-}\bigg]\mathscr{F}_+^L(\mathscr{M_+})
	\bigg\rangle\bigg|_{\textrm{PL}} \, .
\end{equation}
In what follows we particularize for our case at hand, namely  $\mathscr{M}_-^L = \{ L_-^{ij}, I_-^{ij} \}$. Without loss of generality, the angular momentum and quadrupole may be decomposed as
\begin{align}
	L_{-}^{ij} & = 2f_1 \big ( \bx_-^{[i}\bv_+^{j]} + \bx_+^{[i}\bv_-^{j]} \big) \\
		I_-^{ij} & = \left(f_2\,  \bx_-^{(i} \bx_+^{j)} + f_3\, \bx_-^{(i} \bv_+^{j)}+ f_4\, \bv_-^{(i} \bx_+^{j)}+ f_5\, \bv_-^{(i} \bv_+^{j)}\right)  - {\rm traces} \, , 
\end{align}
where the $f_n$'s, $n=1 \cdots 5$, are only functions of the  $+$ variables (with other contributions annihilated by the PL). For instance, at leading order we have
\begin{equation}
	I_-^{ij} = \sum_a m_a \bigg( 2 \bx_{a, -}^{(i}\bx_{a, +}^{j)}- \frac{2}{3}\delta^{ij}\bx_{a, -}\cdot \bx_{a, +} \bigg) \,.
	\label{eq:Iminusexp}
\end{equation}
From the general form in the above expressions we find
\begin{align}
	\bx_{i} \frac{\partial L_-^{ab}}{ \partial \bx^{j}_-} 
	+ \bv_{i} 
	\frac{\partial L_-^{ab}}{ \partial \bv^{j}_-}\bigg|_{\textrm{PL}} & =  
	 f_1\big(
	 \bx_i \bv^b \delta_{j}{}^{a} + \bx^a \bv_i \delta_{j}{}^{b} - (a \leftrightarrow b)	\big)  = L_{i}{}^b\delta_j{}^a - L_i{}^a \delta_j{}^b \, \\
	 	\bx_{i} \frac{\partial I_-^{ab}}{ \partial\bx^{j}_-} 
	+ \bv_{i} 
	\frac{\partial I_-^{ab}}{ \partial \bv^{j}_-}\bigg|_{\textrm{PL}} & =  
	2 \bx_i\big(
	f_2\, \delta_{j}{}^{(a}\bx^{b)} + f_3\, \delta_{j}{}^{(a}\bv^{b)}
	\big) 
	+ 2 \bv_i\big(
	f_4\, \delta_{j}{}^{(a}\bx^{b)} + f_5\, \delta_{j}{}^{(a}\bv^{b)}
	\big) \notag \\
	& = 2 I_{i}{}^{(a}\delta_j{}^{b)} + O\big( \delta_{i}{}^{(a}\delta_j{}^{b)} \big) \, , 
	\label{eq:PDIminusL}
\end{align}
where the last term represent the traces (that ultimately cancel out once contracted with antisymmetric terms). From here  we arrive at the general structure for the flux of angular momentum 
\begin{align}
	\langle\dot{\bL}_{k} \rangle = \varepsilon_{kij} \left\langle I_{i\ell}\Big(
	\mathscr{F}_{\ell j}(\mathscr{M}) + \mathscr{F}_{j \ell}(\mathscr{M})
	\Big) +
	 L_{i\ell} \Big(
	\mathscr{F}_{j \ell}(\mathscr{M}) -  \mathscr{F}_{\ell j}(\mathscr{M}) 
	\Big)\right\rangle \, .
	\label{eq:Ldotfinalc}
\end{align}

\section{Abraham-Lorentz conservative effects} \label{RRQED}

The conservative relativistic scattering of two (nonspinning) charges in electrodynamics due to radiative effects was obtained in~\cite{Bern:2023ccb}, at fourth order in the coupling (see ancillary file in \cite{Bern:2023ccb}). Since (classical) electromagnetism is inherently a linear theory, the existence of such conservative-like terms can only be associated with contributions at second-order in the Abraham-(Dirac)-Lorentz force. We reproduce here this effect, at leading order in the PN expansion, from the point of view of the EFT approach~\cite{Goldberger:2009qd,Galley:2010es}.\vskip 4pt 

At leading order in the multipole expansion, and ignoring the gravitational field, the long-distance effective action for the two-body system takes the form
\begin{equation}
	S_{\rm eff}= -\frac{1}{4}\int \dd^4 x F_{\mu\nu}F^{\mu\nu} - \sqrt{4\pi \alpha} (Q_1+Q_2)\int \dd t\,  A^\mu V_\mu + \sqrt{4\pi \alpha} \int \dd t \, \bd\cdot \bE + \cdots\,,
\end{equation}
\begin{align}
& \alpha\equiv e^2/4\pi	\,, \qquad F_{\mu\nu} = 2 \partial_{[\mu}A_{\nu]} \, , \qquad E_i = - F_{0i}\\
	& \quad V^\mu = \delta^\mu_0\,, \qquad \bd(t) = Q_1 \bx_1(t) + Q_2 \bx_2(t) \,,\notag
\end{align}
in the center-of-mass frame. We have only kept the coupling to the dipole, $\bd$, which is the relevant term at leading PN order. In what follows we denote as $q_a \equiv Q_a/m_a$ ($a=1,2$) the charge/mass ratio. 

\subsubsection*{\textit{In-in} computation}

We follow the same steps as in the gravitational case, yielding
\begin{align}
	i S[\bx_{\pm}] = i (4\pi \alpha) \int \frac{\dd \omega}{2\pi}\int_{\bk}
		\bigg[
		k_i k_j  d^i_-(-\omega)d_+^j(\omega) -\omega^2 d^i_-(-\omega)d_+^i(\omega) 
		\bigg]\frac{1}{(\omega +i0^+)^2-\bk^2} \, ,
\end{align}
which reduces to 
\begin{align}
	S[\bx_\pm] & = \frac{2\alpha}{3} \int \frac{\dd \omega}{2\pi}
		(i\omega^3) d^i_-(-\omega)d_+^i(\omega) 
		 = \frac{2\alpha}{3} \int \dd t \, \bd_-(t)\cdot \bd\ord{3}_+(t) \, .
		\label{eq:SRRinin}
\end{align}

From the in-in action we get the radiation-reaction (Abraham-Lorentz) acceleration for each particle 
\begin{equation}
	 \ba_{a\rm (RR)}= \frac{2\alpha}{3m_a}\frac{\partial \bd_-(t)}{\partial \bx_{a,-}(t)} \cdot \bd\ord{3}_+(t)\bigg|_{\textrm{PL}} = \frac{2\alpha}{3} q_a \bd\ord{3}(t) = \frac{2\alpha}{3} q_a \big( m_1 q_1 \dddot{\bx}_1 + m_2 q_2 \dddot{\bx}_2 \big) \, .\label{alforce}
\end{equation}

In order to obtain the leading order contribution, we replace the acceleration on the right-hand side  of \eqref{alforce} by the Coulomb force, yielding the relative acceleration\footnote{Notice that it vanishes when $q_1=q_2$.}
\begin{subequations}
\begin{align}
	 \ba_{\rm (RR)}& \equiv  \ba_{1\rm (RR)} - \ba_{2\rm (RR)}  = \frac{2}{3}\alpha^2 M \nu \, q_1 q_2 (q_1 - q_2)^2 \frac{\dd}{\dd t} \left(\frac{\br(t)}{r^3(t)}\right) \, . 
\end{align}
\end{subequations}
Since it is a total derivative, the contribution from this acceleration to the total impulse is clearly zero, regardless of the trajectory. That means that the only correction comes from the radiation-reaction deflected trajectory (or `iteration' \cite{Kalin:2020mvi}) $\br(t)  = \bb + \bv_\infty t + \delta^{\rm (RR)} \br(t)$, which solves the equation
\begin{equation}
	\frac{\dd^2}{\dd t^2} \delta\ord{\textrm{RR}}\br(t) = \ba\ord{\textrm{RR}} \, ,
\end{equation}
into the Coulomb force, yielding%
\begin{subequations}
	\begin{align}
		\Delta\ord{\textrm{RR}}\bp_1 & = \frac{\alpha^3 M^4}{b^3}\frac{q_1^2 q_2^2 (q_1 -q_2)^2 \nu}{3 v_\infty^2}\bigg[-4 \frac{\bb}{b} - \pi \frac{\bv_\infty}{v_\infty}\bigg] \, ,\\ 
		\Delta\ord{\textrm{RR}}\bp_2 & = \frac{\alpha^3 M^4}{b^3}\frac{q_1^2 q_2^2 (q_1 -q_2)^2 \nu}{3 v_\infty^2}\bigg[4 \frac{\bb}{b} + \pi \frac{\bv_\infty}{v_\infty}\bigg] \, ,
	\end{align}
\end{subequations}
where, as in the main text, $(\bb,\bv_\infty)$ are the impact parameter and relative velocity at infinity, respectively.  This agrees with the nonrelativistic expansion of the result in~\cite{Kalin:2022hph}.\vskip 4pt

The acceleration at second order in the radiation-reaction (RR$^2$) can be obtained by plugging back the radiation-reaction acceleration onto the right-hand side of the force,  
\begin{align}
		\label{eq:aRR2d4}
	\ba_{a\rm (RR^2)}
	& =  \frac{4}{9}\alpha^2 q_a \big(m_1 q_1^2 + m_2 q_2^2 \big) \frac{\dd}{\dd t} \bigg[
		m_1 q_1\dddot{\bx}_1 + m_2 q_2\dddot{\bx}_2 
		\bigg] \, \\
		& =\frac{4}{9} \alpha^3 M^2 \nu  q_a \big(m_1 q_1^2 + m_2 q_2^2 \big)q_1 q_2 (q_1-q_2) \frac{\dd^2}{\dd t^2} \frac{\br(t)}{r^3(t)}\,, \notag
\end{align}
where in the last equality we inputed the Coulomb acceleration.  Upon evaluating the time derivatives, we find%
\beq
\begin{aligned}
	\ba_{a\rm (RR^2)}
 & =\frac{4}{9} \alpha^3 M^2 \nu  q_a \big(m_1 q_1^2 + m_2 q_2^2 \big)q_1 q_2 (q_1-q_2)\\
	&\times \bigg[
	\bigg( 15 \frac{(\bv\cdot \br)^2}{r^7} - 3 \frac{v^2}{r^5}\bigg) \br
	-6 \frac{(\br\cdot \bv)}{r^5} \bv - 2 \alpha m q_1 q_2 \frac{\br}{r^6}
	\bigg] \, .
	\label{eq:RR2nodt}
\end{aligned}
\eeq
such that
\beq
\begin{aligned}
	\ba_{\rm (RR^2)}
 & =\frac{4}{9} \alpha^3 M^2 \nu \big(m_1 q_1^2 + m_2 q_2^2 \big)q_1 q_2 (q_1-q_2)^2 \\
&\times	\bigg[
	\bigg( 15 \frac{(\bv\cdot \br)^2}{r^7} - 3 \frac{v^2}{r^5}\bigg) \br
	-6 \frac{(\br\cdot \bv)}{r^5} \bv - 2 \alpha m q_1 q_2 \frac{\br}{r^6}
	\bigg] \, ,
	\label{eq:aRR2rel}
\end{aligned}
\eeq
for the relative acceleration, where the $\alpha^4$ term arises from the inclusion, once again, of the Coulomb force. 
Because of the structure of the radiation-reaction-square force, it is straightforward to show that\footnote{Notice that in order to obtain this result from the expression in \eqref{eq:aRR2rel} we must also take into account not only straight motion but also the leading order deflection due to the Coulomb field.}   
\begin{equation}
	\int \dd t \ba_{a}\ord{\textrm{RR}^2} \sim \bigg[ \frac{\dd}{\dd t} \frac{\br(t)}{r^3(t)} \bigg] \bigg|^\infty_{-\infty} = 0 \, .
\end{equation}
Since this integral vanishes, the nontrivial contribution the total impulse comes from evaluating the Coulomb acceleration on the trajectory deflected by the force in \eqref{eq:aRR2rel} to the desired order. We find
\begin{align}
	\Delta^{(\textrm{RR}^2)}\bp_1 = -\pi\frac{\alpha^4 M^4 \nu^2}{b^4}\big(m_1 q_1^2 + m_2 q_2^2 \big)q_1^2 q_2^2 (q_1-q_2)^2\frac{1}{3v_\infty}\frac{\bb}{b}\, .
\end{align}
We discover $\Delta\ord{\textrm{RR}^2}\bp_1 = -\Delta\ord{\textrm{RR}^2}\bp_2$, such that the contribution to the  scattering angle becomes (using the same convention for a positive angle along the $-\hat\bb$ direction) 
\begin{equation}
	\chi^{\rm tot}_{(\textrm{RR}^2)} = \frac{|\Delta\ord{\textrm{RR}^2}\bp_1|}{M\nu v_\infty} = 
	\pi\frac{\alpha^4 M^3 \nu}{3 b^4 v_\infty^2}\big(m_1 q_1^2 + m_2 q_2^2 \big)q_1^2 q_2^2 (q_1-q_2)^2 \,,
	\label{eq:chiRR2em}
\end{equation}
which, amusingly, also agrees at leading PN order with the conservative result in~\cite{Bern:2023ccb}. As we demonstrate next, this is expected due to the nature of the radiation-reaction-square force, and will be recovered as well from the in-out (Feynman) computation, as we demonstrate momentarily. 

\subsubsection*{\textit{RR$^2$ force is conservative}} 
\
Before proceeding, let us demonstrate that the radiation-reaction-square force in \eqref{eq:RR2nodt}, to the order we are concerned about, is conservative. We already found the conservation of momentum.  Let us look now at the loss of energy of each body, say particle 1, 
\begin{align}
	\dot E_1 \equiv m_1 \bv_1 \cdot \ba_{1(\textrm{RR}^2)} = A m_1 q_1 \dot{\bx}_1 \cdot \frac{\dd}{\dd t}\bigg[
		m_1 q_1\dddot{\bx}_1 + m_2 q_2\dddot{\bx}_2
		\bigg] \, ,
\end{align}
where, in order to simplify notation, we introduced $A \equiv (4/9)\alpha^2 \big(m_1 q_1^2 + m_2 q_2^2 \big)$. After performing the following manipulations
\begin{align}
	\dot{\bx}_1 \cdot \frac{\dd}{\dd t} \dddot{\bx}_1 & = -\ddot{\bx}_1 \cdot \dddot{\bx}_1 + \frac{\dd}{\dd t} (\cdots ) = - \frac{1}{2}\frac{\dd}{\dd t} \Big(
	\ddot{\bx}_1^2 + \cdots 
	\Big) \, ,\\ 
	\dot{\bx}_1 \cdot \frac{\dd}{\dd t} \dddot{\bx}_2 & = -\ddot{\bx}_1 \cdot \dddot{\bx}_2 + \frac{\dd}{\dd t} (\cdots )  \, .
\end{align}
we find, up to total time derivatives,
\begin{equation}
	\dot E_1 = - A m_1  m_2 q_1 q_2\, \ddot{\bx}_1 \cdot \dddot{\bx}_2 + \frac{\dd}{\dd t} (\cdots ) \, .
\end{equation}
and similarly,
\begin{equation}
	\dot E_2 = - A m_2  m_1 q_1 q_2 \ddot{\bx}_2 \cdot \dddot{\bx}_1 + \frac{\dd}{\dd t} (\cdots ) \, .
\end{equation}
so that
\begin{align}
	\dot E_1 + \dot E_2 & = - A m_2 q_2 m_1 q_1 \Big(
	\ddot{\bx}_1 \cdot \dddot{\bx}_2
	+ \ddot{\bx}_2 \cdot \dddot{\bx}_1  
	\Big) + \frac{\dd}{\dd t} (\cdots ) \notag \\
	& = - A m_2 q_2 m_1 q_1 \frac{\dd}{\dd t} \Big(
	\ddot{\bx}_1 \cdot \ddot{\bx}_2 + \cdots
	\Big) =  \frac{\dd}{\dd t} ( \cdots )\,, 
\end{align}
becomes a total derivative, which  can then be moved to the left-hand side to define a new conserved quantity. Similar steps can be performed for the angular momentum. In that case we have
\begin{align}
	\dot \bL^i_{1} & = m_1 \epsilon^{ijk} \bx_1^j \ba^k_{1(\rm RR^2)}  = A m_1 q_1 \epsilon^{ijk} \bx_1^j \frac{\dd}{\dd t}\bigg[
		m_1 q_1\dddot{\bx}^k_1 + m_2 q_2\dddot{\bx}^k_2
		\bigg] \, ,
\end{align}
and using the identities,
\begin{align}
	\bx_1^j \frac{\dd}{\dd t}
		\dddot{\bx}^k_1  & = - \dot{\bx}_1^j \dddot{\bx}^k_1 + \frac{\dd}{\dd t} ( \cdots ) = \ddot{\bx}_1^j \ddot{\bx}^k_1 + \frac{\dd}{\dd t} ( \cdots )  \, . 
		\label{eq:symjk}\\
	\bx_1^j \frac{\dd}{\dd t}
		\dddot{\bx}^k_2  & = - \dot{\bx}_1^j \dddot{x}^k_2 + \frac{\dd}{\dd t} ( \cdots ) = \ddot{\bx}_1^j \ddot{\bx}^k_2 + \frac{\dd}{\dd t} ( \cdots )  \, ,
\end{align}
the only nontrivial contribution becomes
\begin{equation}
	\dot\bL^i_1 = A m_1 q_1 m_2 q_2 \epsilon^{ijk} \ddot{\bx}_1^j \ddot{\bx}^k_2 + \frac{\dd}{\dd t} ( \cdots ) \, .
\end{equation}
and likewise for particle 2,%
\begin{equation}
	\dot\bL^i_2 = A m_2 q_2 m_1 q_1 \epsilon^{ijk} \ddot{\bx}_2^j \ddot{\bx}^k_1 + \frac{\dd}{\dd t} ( \cdots ) \, , 
\end{equation}
such that, for the total loss of angular momentum,
\begin{equation}
	\dot\bL^i_1+\dot \bL_2^i = A m_2  m_1 q_2 q_1 \epsilon^{ijk} \Big(
	\ddot{\bx}_1^j \ddot{\bx}^k_2 + 
	\ddot{\bx}_2^j \ddot{\bx}^k_1
	\Big)  + \frac{\dd}{\dd t} ( \cdots ) =  \frac{\dd}{\dd t} ( \cdots )  \, ,
\end{equation}
which, once again, becomes a total derivative. 

\subsection*{\textit{In-out} computation}

We  show now how to derive the conservative-like contribution to the scattering angle using the in-out approach.  The effective action is obtained by integrating out the electromagnetic field. Following similar steps as before, we find%
\beq
\begin{aligned}
	S^{\rm cons}_{\textrm{(RR)}} & = -\frac{1}{3}(4\pi \alpha) \int \frac{\dd \omega}{2\pi}
		\omega^2 d^i(-\omega)d^i(\omega) 
		\int_{\bk} \frac{1}{(\omega +i0^+)^2-\bk^2} \label{eq:SRRinout}
 \\
		& = \frac{1}{3}\alpha \int \frac{\dd \omega}{2\pi}
		(i\omega^3) \sg(\omega) d^i(-\omega)d^i(\omega) 
		 = -i\frac{\alpha}{3\pi} \int \dd t_1 \dd t_2 \frac{\PV}{t_1 - t_2} \bd(t_1) \cdot \bd\ord{3}(t_2) \, ,
\end{aligned}
\eeq
The overall factor of $1/2$ difference with respect to the in-in result in~\eqref{eq:SRRinin} is due to the symmetry of the diagram, and in the last step we used \eqref{absw}. The reader will immediately notice that, provided the time derivatives of the dipole term are real functions of time (as expected from Coulomb-like interactions) the real part of the above in-out effective action vanishes, which is consistent with the fact that there is no conservative contribution at leading order in the radiation-reaction force. (The imaginary part, on the other hand, directly leads to the well-known dipole emission formula: $P \sim \bd^{(2)}\cdot \bd^{(2)}$, as expected from the optical theorem~\cite{Porto:2016pyg}.) We can, nonetheless, obtain an acceleration using the generalized Euler-Lagrangian equations \cite{Galley:2014wla}, 
\begin{equation}
	\bF_a  = \frac{\partial \mathcal{L}}{\partial \bx_a(t)}
	- \frac{\dd}{\dd t}\frac{\partial \mathcal{L}}{\partial \dot{\bx}_a(t)}
	+ \frac{\dd^2}{\dd t^2}\frac{\partial \mathcal{L}}{\partial \ddot{\bx}_a(t)}
	- \frac{\dd^3}{\dd t^3}\frac{\partial \mathcal{L}}{\partial \dddot{\bx}_a(t)} + \cdots \,,
\end{equation}
and we get
\begin{align}
	\bF^{\rm cons}_{\rm (RR)a} & = -\frac{i\alpha}{3\pi} Q_a \int \dd t_1 \dd t_1 \bigg[
	\delta(t-t_1)\frac{\PV}{t_1-t_2} \bd\ord{3}(t_2)  - \bd(t_1) \frac{\dd^3}{\dd t^3}\bigg( \frac{\PV}{t_1-t_2} \delta(t-t_2) \bigg)
	\bigg] \, 	\label{eq:FRR} \\
	& = -\frac{2i\alpha}{3\pi}Q_a \int \dd t' \frac{\PV}{t-t'} \bd\ord{3}(t') = -\frac{2i\alpha}{3\pi} m_a q_a \int \dd t' \frac{\PV}{t-t'}\big(
	 m_1 q_1 \dddot{\bx}_1(t') + m_2 q_2 \dddot{\bx}_2(t')
	\big) \, .
\notag\end{align}
The fact that the leading radiation-reaction force is imaginary is a consequence of the fact that radiative effects dissipative energy at leading order, and therefore, from the decomposition of the retarded Green's function into Feynman plus a {\it reactive} term \cite{Kalin:2022hph}, only the latter contributes. However, that is no longer the case at second order, as we demonstrate in what follows.\vskip 4pt

To obtain the radiation-reaction-square effects, we replace the second derivatives in~\eqref{eq:FRR} with the very same acceleration, yielding
\begin{align}
	\ba^{\rm cons}_{{\rm (RR^2)}a} &  = -\frac{4\alpha^2}{9\pi^2} q_a \int \dd t' \frac{\PV}{t-t'}
	\bigg[
	 m_1 q_1^2 \frac{\dd}{\dd t'}\int \dd t'' \frac{\PV}{t'-t''}
	 \bigg( m_1 q_1\dddot{\bx}_1(t'') + m_2 q_2 \dddot{\bx}_2(t'')
	  \bigg) \notag \\
	  & \hspace{3cm} + m_2 q_2^2 \frac{\dd}{\dd t'}\int \dd t'' \frac{\PV}{t'-t''}
	 \bigg( m_1 q_1\dddot{\bx}_1(t'') + m_2 q_2 \dddot{\bx}_2(t'')
	  \bigg)
	\bigg] \notag \\
	& = -\frac{4\alpha^2}{9\pi^2} q_a (m_1 q_1^2 + m_2 q_2^2) \int \dd t' \frac{\PV}{t-t'} \int \dd t'' \frac{\PV}{t'-t''} \bigg( m_1 q_1\bx\ord{4}_1(t'') + m_2 q_2 \bx\ord{4}_2(t'') 
	  \bigg) \, .
\end{align}
We invoke again the Poincar\'e-Bertrand theorem \cite{Davies:1990fe,Davies:1996gee} (see \eqref{pbthm}) and exchange the order of integration, yielding
\begin{align}
	\ba^{\rm cons}_{{\rm (RR^2)}a}&  = -\frac{4\alpha^2}{9\pi^2} q_a (m_1 q_1^2 + m_2 q_2^2) \int \dd t'' \big( m_1 q_1\bx\ord{4}_1(t'') + m_2 q_2 \bx\ord{4}_2(t'')
	  \big) \int \dd t' \frac{\PV}{(t-t')(t'-t'')} \notag \\
	  &  +\frac{4\alpha^2}{9} q_a (m_1 q_1^2 + m_2 q_2^2)  \big( m_1 q_1\bx\ord{4}_1(t) + m_2 q_2 \bx\ord{4}_2(t) \big)\notag \\
	  &= \frac{4\alpha^2}{9} q_a (m_1 q_1^2 + m_2 q_2^2)  \big( m_1 q_1\bx\ord{4}_1(t) + m_2 q_2 \bx\ord{4}_2(t) \big)\,,

\end{align}
 where in the last equality we used the fact that \cite{Davies:1990fe,Davies:1996gee}
 \beq 
 \int \dd t \frac{\PV}{(t-t')(t'-t'')} = 0\,.\label{pvzero}
 \eeq
We are thus left with the exact same acceleration in~\eqref{eq:aRR2d4} from the in-in approach, which implies that the contributions from the reactive terms all vanish at this order. Following the same steps as before, we get the same value for the scattering angle, consistently with the relativistic result in \cite{Bern:2023ccb}.

\subsubsection*{\textit{Radial action} }

It is instructive to derive the contribution from radiation-reaction-square effects also directly at the level of the (radial) action. This can be done replacing the second derivative in~\eqref{eq:SRRinout} with the acceleration in~\eqref{eq:FRR}, together with \eqref{pvzero}, yielding
\begin{align}
	S_{(\textrm{RR}^2)}^{\rm cons} & =-\frac{2\alpha^2}{9\pi^2}
	\big(
	m_1 q_1^2 + m_2 q_2^2
	\big) \int \dd t_1  \bd(t_1) \cdot \int \dd t_2 \frac{\PV}{t_1 - t_2} \int \dd t_3 \frac{\PV}{t_2 - t_3}\bd_{\rm (RR)}\ord{4}(t_3) \notag \\
	& = \frac{2\alpha^2}{9}
	\big(
	m_1 q_1^2 + m_2 q_2^2
	\big) \int \dd t \,\bd(t) \cdot \bd_{\rm (RR)}\ord{4}(t) \,,
	\label{eq:RR2exp}
\end{align}
where
\begin{equation}
	\bd_{\rm (RR)}\ord{4}(t) = \alpha M^2 \nu q_1 q_2 (q_1 - q_2) \frac{\dd^2}{\dd t^2}\left( \frac{\br(t)}{r^3(t)} \right)\, ,
\end{equation}
such that
\begin{equation}
	S_{(\textrm{RR}^2)}^{\rm cons}  = \frac{2\alpha^3}{9} M^2 \nu
	\big(
	m_1 q_1^2 + m_2 q_2^2
	\big)  q_1 q_2(q_1 - q_2) \int \dd t \,\bd(t)\cdot \frac{\dd^2}{\dd t^2} \left(\frac{\br(t)}{r^3(t)}\right) \, .
\end{equation}
From here we can relate the conservative action, evaluated on the (leading) Coulomb trajectory, to the scattering angle via (see e.g. \cite{Bini:2021gat})
\beq
\chi^{\rm cons}_{\rm (RR^2)} = -\frac{\partial}{\partial J}S_{(\textrm{RR}^2)}^{\rm cons}\,, 
\eeq
with $J = M\nu v_\infty b$ the total angular momentum. Upon inputing the solution for the trajectory, we arrive at
\beq
S_{(\textrm{RR}^2)}^{\rm cons} = \pi\frac{\alpha^4 M^4 \nu^2}{9 b^3 v_\infty}\big(m_1 q_1^2 + m_2 q_2^2 \big)q_1^2 q_2^2 (q_1-q_2)^2 \, ,
\eeq
from which we find
\begin{equation}
	\chi^{\rm cons}_{\rm (RR^2)}=\pi\frac{\alpha^4 M^3 \nu}{3 b^4 v_\infty^2}\big(m_1 q_1^2 + m_2 q_2^2 \big)q_1^2 q_2^2 (q_1-q_2)^2 \,,
\end{equation}
in agreement with~\eqref{eq:chiRR2em}, which implies as before that $\chi^{\rm tot}_{\rm (RR^2)}=\chi^{\rm cons}_{\rm (RR^2)}$ at this order. 
\bibliographystyle{JHEP}
\bibliography{refmem5PN}

\end{document}

%% file: diags.tex
\usepackage{tikz}
\usepackage[compat=1.1.0]{tikz-feynman}

\newcommand{\gravhb}{
\begin{tikzpicture}[baseline]
\begin{feynman}
\vertex (a);
\vertex [right=1cm of a] (b);
\diagram* {
(a) -- [plain] (b)
}; 
\end{feynman}
\end{tikzpicture}
}

\newcommand{\gravh}{
\begin{tikzpicture}[baseline]
\begin{feynman}
\vertex (a);
\vertex [right=1cm of a] (b);
\diagram* {
(a) -- [photon] (b)
}; 
\end{feynman}
\end{tikzpicture}
}

\newcommand{\Csoure}{
\begin{tikzpicture}[baseline]
\begin{feynman}
\vertex [dot, minimum size=0.25cm, label=180:$\mathscr{M}_\pm$] (a) {};
\diagram* {
}; 
\end{feynman}
\end{tikzpicture}
}

\newcommand{\retg}{
\begin{tikzpicture}[baseline]
\begin{feynman}
\vertex [label=90:$\scriptstyle{\mu\nu}$](a);
\vertex [right=0.85cm of a] (a');
\vertex [right=1.5cm of a, label=90:$\scriptstyle{\rho\sigma}$] (b);
\diagram* {
(a) -- [scalmom, momentum={[arrow shorten=0.3pt]$k$}] (b)
}; 
\end{feynman}
\end{tikzpicture}
}

\newcommand{\advg}{
\begin{tikzpicture}[baseline]
\begin{feynman}
\vertex [label=90:$\scriptstyle{\mu\nu}$](a);
\vertex [right=0.85cm of a] (a');
\vertex [right=1.5cm of a, label=90:$\scriptstyle{\rho\sigma}$] (b);
\diagram* {
(a) -- [antiscalmom, momentum={[arrow shorten=0.3pt]$k$}] (b)
}; 
\end{feynman}
\end{tikzpicture}
}

\newcommand{\cubicbackpp}{
\begin{tikzpicture}[baseline]
\begin{feynman}
\vertex (a);
\vertex [right=1.2cm of a] (b);
\vertex [right=1.2cm of b,label=90:$\scriptstyle{h^{\mu\nu}_-}$] (c);
\vertex [above=0.7cm of a, label=90:$\scriptstyle{(\alpha_1\beta_1, +)}$] (du);
\vertex [below=0.7cm of a, label=270:$\scriptstyle{(\alpha_2\beta_2, +)}$] (dd);
\diagram* {
(b) -- [photon, momentum'={[arrow shorten=0.3pt]$k$}] (c),
(du) -- [scalmom, momentum={[arrow shorten=0.3pt]$k_1$}] (b),
(dd) -- [scalmom, momentum'={[arrow shorten=0.3pt]$k_2$}] (b)
}; 
\end{feynman}
\end{tikzpicture}
}

\newcommand{\Mphb}{
\begin{tikzpicture}[baseline]
\begin{feynman}
\vertex [dot, minimum size=0.25cm, label=180:$\{M_+\text{, } L_+\text{, } I_+\}$] (a) {};
\vertex [right=1cm of a] (d);
\vertex [right=0.2cm of d, label=90:$\scriptstyle{h^{\mu\nu}_{-}}$] (f);
\diagram* {
(a) -- [photon, momentum={[arrow shorten=0.3pt]$k$}] (d)
}; 
\end{feynman}
\end{tikzpicture}
}

\newcommand{\Mph}{
\begin{tikzpicture}[baseline]
\begin{feynman}
\vertex [dot, minimum size=0.25cm, label=180:$\{M_+\text{, } L_+\text{, } I_+\}$] (a) {};
\vertex [right=1cm of a] (d);
\diagram* {
(a) -- [scalmom, momentum={[arrow shorten=0.3pt]$k$}] (d)
}; 
\end{feynman}
\end{tikzpicture}
}

\newcommand{\Mmh}{
\begin{tikzpicture}[baseline]
\begin{feynman}
\vertex [dot, minimum size=0.25cm, label=180:$\{M_-\text{, } L_-\text{, } I_-\}$] (a) {};
\vertex [right=1cm of a] (d);
\diagram* {
(a) -- [antiscalmom, momentum={[arrow shorten=0.3pt]$k$}] (d)
}; 
\end{feynman}
\end{tikzpicture}
}

\newcommand{\Mphbtwo}{
\begin{tikzpicture}[baseline]
\begin{feynman}
\vertex [dot, minimum size=0.25cm, label=180:$\{M_+\text{, } L_+\text{, } I_+\}$] (a) {};
\vertex [right=1cm of a] (d);
\vertex [above=0.6cm of d, label=90:$\scriptstyle{h^{\mu\nu}_{-}}$] (i);
\vertex [below=0.6cm of d] (f);
\diagram* {
(a) -- [photon, momentum={[arrow shorten=0.3pt]$k$}] (i),
(a) -- [antiscalmom, momentum'={[arrow shorten=0.3pt]$q$}] (f)
}; 
\end{feynman}
\end{tikzpicture}
}

\newcommand{\Mmhtwo}{
\begin{tikzpicture}[baseline]
\begin{feynman}
\vertex [dot, minimum size=0.25cm, label=180:$\{M_-\text{, } L_-\text{, } I_-\}$] (a) {};
\vertex [right=1cm of a] (d);
\vertex [above=0.6cm of d, label=90:$\scriptstyle{h^{\mu\nu}_{-}}$] (i);
\vertex [below=0.6cm of d] (f);
\diagram* {
(a) -- [photon, momentum={[arrow shorten=0.3pt]$k$}] (i),
(a) -- [scalmom, momentum'={[arrow shorten=0.3pt]$q$}] (f)
}; 
\end{feynman}
\end{tikzpicture}
}

\newcommand{\Tmempom}{
\begin{tikzpicture}[baseline]
\begin{feynman}
\vertex [dot, minimum size=0.25cm, label=270:${\scriptstyle I_+(\omega_1)}$] (a) {};
\vertex [right=1cm of a, dot, minimum size=0.25cm, label=270:${\scriptstyle I_+(\omega_2)}$] (c) {};
\vertex [above=1cm of c] (b);
\vertex [right=0.6cm of b] (up);
\vertex [below=0.2cm of b] (b1);
\vertex [left=0.65cm of b1] (b2);
\vertex [above=0.6cm of c] (aux1);
\vertex [right=0.03cm of aux1] (aux2);
\vertex [right=0.34cm of b] (aux3);
\vertex [above=0.03cm of aux3] (aux4);
\diagram* {
(a) -- [photon, out=90, in=180] (b), 
(b) -- [photon, momentum={[arrow shorten=0.2pt]${\scriptstyle (\omega_1+\omega_2\text{,}\bk)}$}] (up), 
(c) -- [photon] (b), 
(a) -- [plain, draw opacity=0, -{Stealth[width=1.5mm,length=1.5mm]},out=90, in=210] (b2),
(c) -- [plain, draw opacity=0, -{Stealth[width=1.5mm,length=1.5mm]}] (aux2),
(b) -- [plain, draw opacity=0, -{Stealth[width=1.5mm,length=1.5mm]}] (aux4)
};
\end{feynman}
\end{tikzpicture}
}

\newcommand{\Tzeromeqom}{
\begin{tikzpicture}[baseline]
\begin{feynman}
\vertex (a) ;
\vertex [right=1cm of a, dot, minimum size=0.25cm, label=270:${\scriptstyle I_-(\omega_3)}$] (c) {};
\vertex [above=1cm of a] (b);
\vertex [below=0.3cm of b] (b1);
\vertex [right=0.79cm of b1] (b2);
\diagram* {
(c) -- [photon, out=90, in=0, momentum'={[arrow shorten=0.3pt]${\scriptstyle (\omega_3\text{,}\bk)}$}] (b), 
(b) -- [plain, draw opacity=0, -{Stealth[width=1.5mm,length=1.5mm]},out=0, in=110] (b2)
};
\end{feynman}
\end{tikzpicture}
}